\documentclass[iop,apjl,revtex4]{emulateapj}
\usepackage{natbib,color}
\usepackage[yyyymmdd]{datetime}
\usepackage[breaklinks,colorlinks,citecolor=blue]{hyperref}

\slugcomment{Draft version \today~ \currenttime}

\shorttitle{7 January 2014 CME Propagation and Geomagnetic Non-Event}
\shortauthors{Mays et al.}

\begin{document}

\title{Propagation of the 7 January 2014 CME and Resulting Geomagnetic Non-Event}

\author{M. L. Mays\altaffilmark{1,2}}

\author{B. J. Thompson\altaffilmark{2}}

\author{L. K. Jian\altaffilmark{3,2}}

\author{R. C. Colaninno\altaffilmark{4}}

\author{D. Odstrcil\altaffilmark{6}}

\author{C. M\"{o}stl\altaffilmark{7,8}}

\author{\\M. Temmer\altaffilmark{8}}

\author{N. P. Savani\altaffilmark{5,2}}

\author{G. Collinson\altaffilmark{1,2}}

\author{A. Taktakishvili\altaffilmark{1,2}}

\author{P. J. MacNeice\altaffilmark{2}}

\author{Y. Zheng\altaffilmark{2}}

\altaffiltext{1}{Catholic University of America, Washington, DC, USA}
\altaffiltext{2}{Heliophysics Science Division, NASA Goddard Space Flight Center, Greenbelt, MD, USA}
\altaffiltext{3}{Department of Astronomy, University of Maryland, College Park, MD, USA}
\altaffiltext{4}{Space Science Division, Naval Research Laboratory, Washington, DC, USA}
\altaffiltext{5}{Solar Section, Applied Physics Laboratory Johns Hopkins, Laurel, MD, USA}
\altaffiltext{6}{George Mason University, Fairfax, VA, USA}
\altaffiltext{7}{Space Research Institute, Austrian Academy of Sciences, Graz, Austria}
\altaffiltext{8}{IGAM-Kanzelh\"{o}he Observatory, Institute of Physics, University of Graz, Graz, Austria}

\begin{abstract}
On 7 January 2014 an X1.2 flare and CME with a radial speed $\approx$\,2500 km\,s$^{-1}$ was observed from near an active region close to disk center. This led many forecasters to estimate a rapid arrival at Earth ($\approx$\,36 hours) and predict a strong geomagnetic storm. However, only a glancing CME arrival was observed at Earth with a transit time of $\approx$\,49 hours and a $K_{\rm P}$ geomagnetic index of only $3-$. We study the interplanetary propagation of this CME using the ensemble Wang--Sheeley--Arge (WSA)--ENLIL+Cone model, that allows a sampling of CME parameter uncertainties.  We explore a series of simulations to isolate the effects of the background solar wind solution, CME shape, tilt, location, size, and speed, and the results are compared with observed {\it in-situ} arrivals at Venus, Earth, and Mars. Our results show that a tilted ellipsoid CME shape improves the initial real-time prediction to better reflect the observed {\it in-situ} signatures and the geomagnetic storm strength.    CME parameters from the Graduated Cylindrical Shell model used as input to WSA--ENLIL+Cone, along with a tilted ellipsoid cloud shape, improve the arrival-time error by 14.5, 18.7, 23.4 hours for Venus, Earth, and Mars respectively.   These results highlight that CME orientation and directionality with respect to observatories play an important role in understanding the propagation of this CME, and for forecasting other glancing CME arrivals.  This study also demonstrates the importance of three-dimensional CME fitting made possible by multiple viewpoint imaging. 
\end{abstract}


\section{Introduction}\label{intro}

A very fast CME ($\approx$\,2500 km\,s$^{-1}$) on 7 January 2014 was associated with an active region near disk center, leading most forecasters world--wide to predict rapid arrivals at Earth with strong geomagnetic impacts on the {\it CME Scoreboard} (\href{http://kauai.ccmc.gsfc.nasa.gov/CMEscoreboard}{{\sf kauai.ccmc.gsfc.nasa.gov/CMEscoreboard}}). The Community Coordinated Modeling Center (CCMC) (located at NASA Goddard Space Flight Center) serves the {\it CME Scoreboard} website to the research community who may submit CME shock arrival-time predictions in real-time using different forecasting methods; this facilitates model validation under real-time conditions.  Users submitted fifteen arrival-time predictions to the {\it CME Scoreboard} for this event, with an average predicted arrival-time of 9 January at 6:35 UT ($\approx$\,36 hours transit time) and average predicted $K_{\rm P}$ range of 6--7.6.   The actual CME arrival-time was at 19:39 UT ($\approx$\,49 hours transit time) and the geomagnetic $K_{\rm P}$ index was $3-$ (2.7). 

\citet{mostl2015} studied the non-radial motion of this CME by $\approx$\,37$^{\circ}\pm$10$^{\circ}$ away from the source region longitude using EUV and coronagraph observations together with {\it in-situ} signatures at Earth and Mars.  They demonstrate that the CME was ``channeled'' by strong nearby active region magnetic fields and open coronal fields into a non-radial propagation direction within 2.1 solar radii (R$_{\odot}$), in contrast to deflection in interplanetary space. \citet{torok2005} and \citet{liu2008} show that instabilities in a modeled magnetic flux rope can lead to a CME or a failed eruption depending on how the overlying coronal field decreases with height.  In a recent study by \citet{thalmann2015} it was shown, using reconstructions of the three-dimensional (3D) coronal magnetic field from a nonlinear force free model, that the overlying magnetic field plays an important role in producing CMEs in the lower corona.  Together with the result from \citet{mostl2015}, this indicates that the coronal magnetic field is an important parameter for early CME characteristics.   

In this work we consider the relative importance of CME geometry with respect to observatories in CME propagation modeling.  Whether a CME will arrive at Earth, its arrival-time, and geoeffectiveness is largely determined by the angle between the CME central axis with respect to Earth and the CME size.  This is complicated by the large uncertainties in CME size, and the related properties of speed and direction.  Many CME propagation models do not take into account the full 3D CME geometry, such as tilt, asymmetric size, or propagation direction out of the ecliptic plane (2D models).  \citet{shen2014} found that for the front-sided full halo CMEs which arrived at Earth in their sample, 74\% of them satisfied the criterion that the CME half angular width was greater than the angle between the CME propagation direction and Sun-Earth line.  Most geoeffective halo CMEs originate from near the central meridian, but only about half of Earth-directed halo CMEs are geoeffective \citep{schwenn2005,wang2002,kim2008}. \citet{moon2005} show that for a sub-selected sample of twelve front-sided halo CMEs the CME direction is an important parameter for determining the geoeffectiveness, as measured by the $Dst$ index.  In their study, a ``direction parameter'', determined from how asymmetric the halo CME appeared in coronagraph observations was well-correlated with $Dst$, while the source region location and CME speed, which is highly affected by projection effects, were not as well-correlated. \citet{kim2008} confirmed this result for a larger sample of 486 front-sided halo CMEs.  

\citet{jian2006} categorized CME total pressure profiles can into three regimes depending on the impact angle with respect to the CME core; the impact parameter affects the observed magnetic field structures and any geomagnetic response. \citet{savani2015a} present a method to predict the CME's magnetic field at Earth based on the \citet{bothmer1994,bothmer1998} helicity rule and the impact parameter relative to the CME flux-rope axis.  They highlight the importance of triangulating CME geometry when predicting the resulting {\it in-situ} properties at Earth.   

Uncertainties in CME properties can be reduced, but not eliminated, by CME reconstruction techniques which take advantage of multiple viewpoint imaging.  \citet{mierla2010} evaluate different techniques to reconstruct the 3D configuration of CMEs from coronagraph data from multiple viewpoints and estimate that the CME propagation direction can be determined to within 10$^{\circ}$. \citet{colaninno2013} show that CME arrival-time predictions based on 3D height-time measurements of {\it Solar TErrestrial RElations Observatory} \citep[\textit{STEREO:}][]{stereo} {\it Sun Earth Connection Coronal and Heliospheric Investigation} \citep[SECCHI:][]{secchi} heliospheric images show a half-day improvement compared with methods using one coronagraph viewpoint. \citet{mostl2014} also find that predicting CME speeds and arrival-times using techniques applied to J-maps constructed from heliospheric images yielded more accurate results than using plane-of-sky coronagraph measurements.   \citet{shi2015} emphasize the importance of using stereoscopy to determine CME parameters for the accuracy of arrival-time for 21 Earth-directed CMEs with the Drag Based Model (DBM: \citet{vrsnak2004,vrsnak2013}) model.  

In order to further understand the interplanetary propagation of the 7 January 2014 CME we use the ensemble WSA--ENLIL+Cone model  implemented at the CCMC. The Wang--Sheeley--Arge (WSA) coronal model \citep{arge2000,arge2004} is used as input to the ENLIL \citep{odstrcil2003} 3D magnetohydrodynamic (MHD) code that computes the background solar wind into which a hydrodynamic ``CME'' cloud is launched.   Ensemble modeling is a collection of forecasts for the same event that represent possible scenarios given the uncertainties associated with the forecasting (such as initial conditions, techniques, and models).  In addition to giving an average forecast, probabilities of events can be determined from the possible scenarios.  In this study we evaluate the sensitivity of the WSA--ENLIL+Cone modeled CME arrival-times to the uncertainties associated with the initial CME parameters of location, speed, size, and shape.

This paper is organized as follows: In Section \ref{obs} we discuss the observations and methods for analyzing CME kinematics to be used as input to the model. The WSA--ENLIL+Cone ensemble model is described in Section \ref{prop}. A series of ensemble CME propagation modeling results are presented in Sections \ref{ensa}--\ref{ensb}, and further compared in Section \ref{compare}.  Finally, we present a summary of the results in Section \ref{disc}.

\section{Observations}\label{obs}
\begin{figure*}
\epsscale{0.49}
\plotone{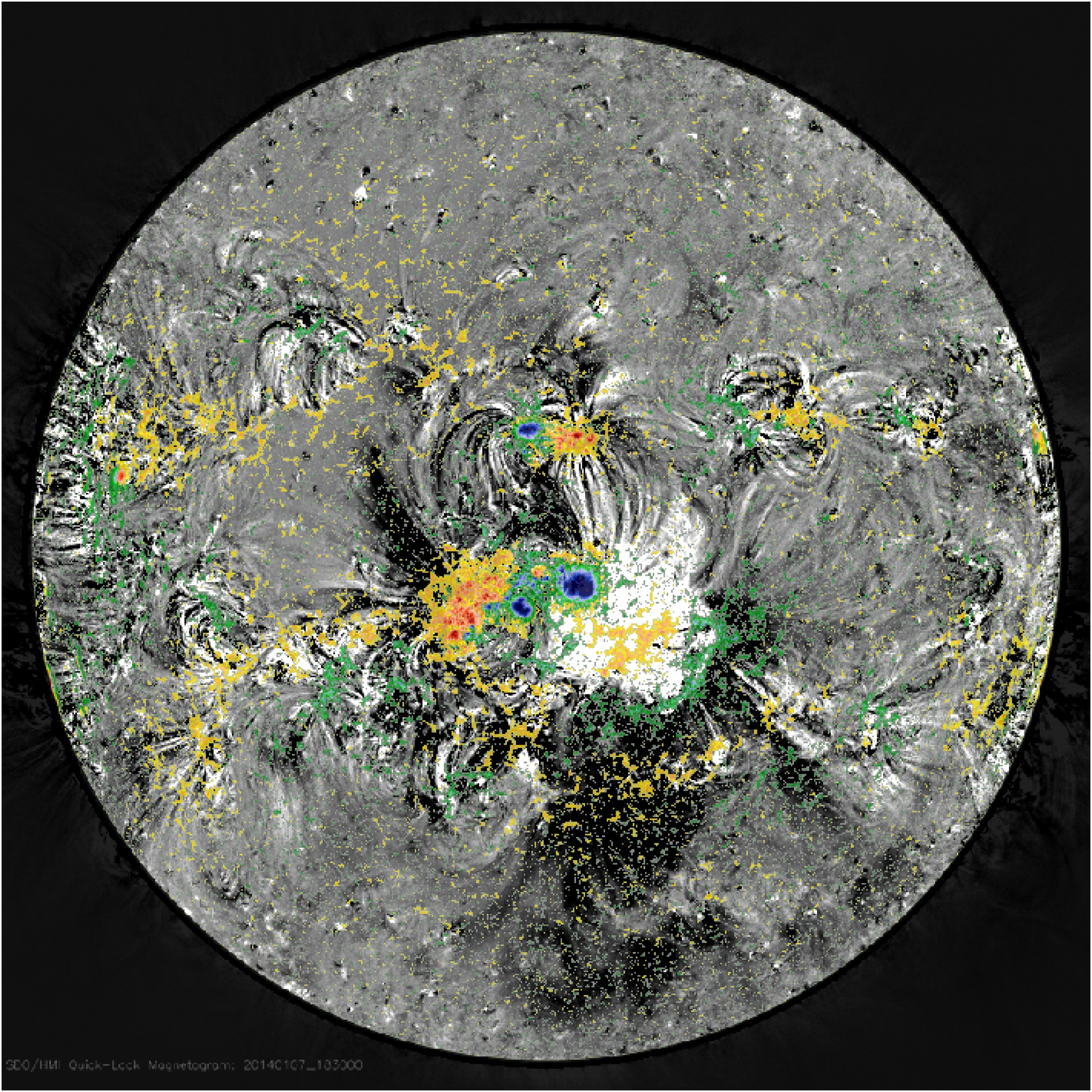}\plotone{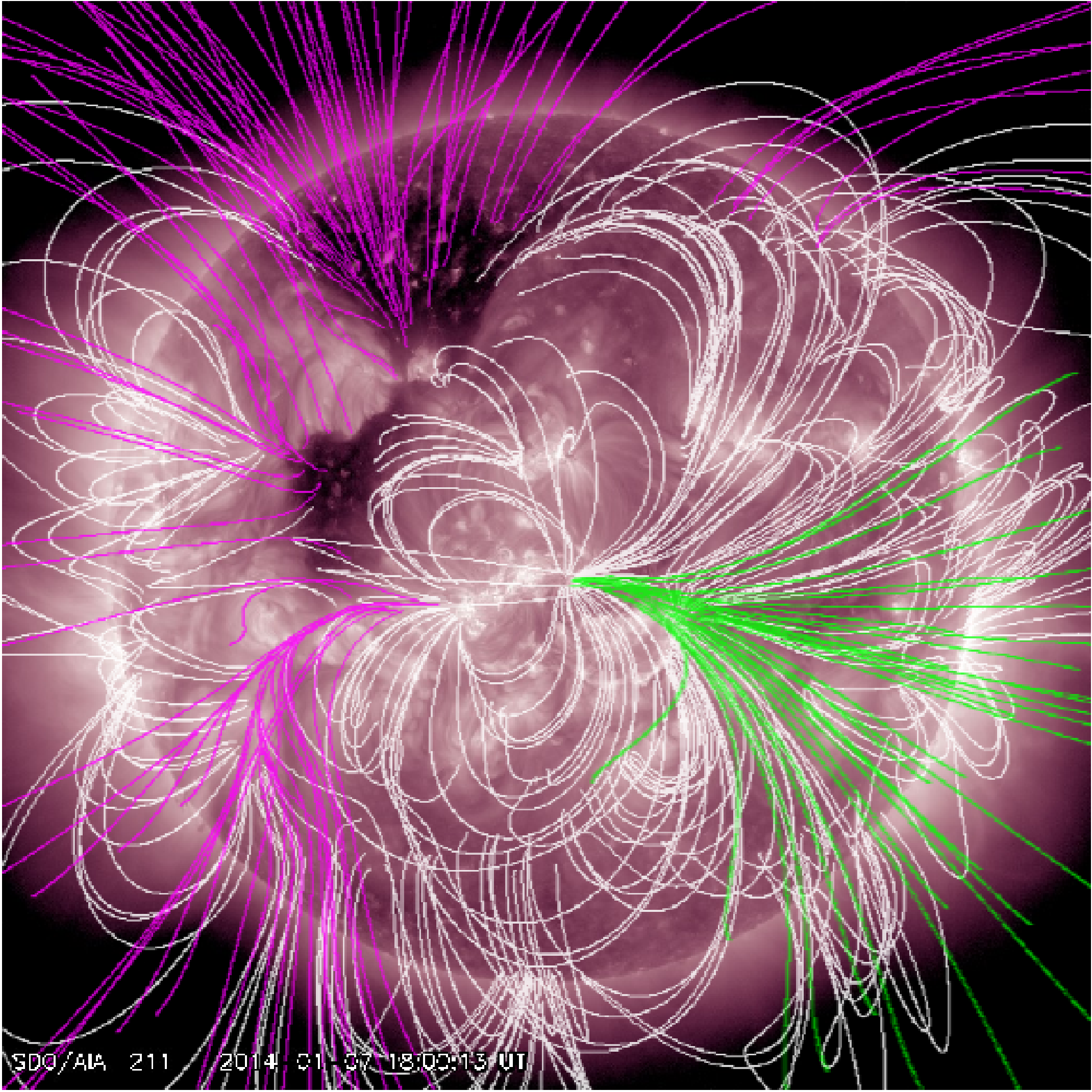}
\caption{(a) SDO/HMI colorized magnetogram image superimposed on a SDO/AIA 193\,\AA\, difference image at 2014-01-07 19:00 UT from which the 18:00 UT image is subtracted (prior to flare).  In the colorized magnetogram overlay, yellow-red colors indicate negative polarity, and green-blue colors indicate positive polarity. Bright post-eruption loops were seen near the active region with an extensive dimming region visible to the south and southwest of the active region with weak negative polarity. (b) SDO/AIA 211\AA~image near 18:00 UT with open (pink: negative polarity, green: positive polarity) and closed (white) magnetic field lines from a PFSS model overlaid.  Two large coronal holes are visible to the northeast of the active region, and open coronal field is present to the west of the region.\label{aia}}
\end{figure*}

The 7 January 2014 CME was associated with an X1.2 class solar flare from between active regions numbered 11943 and 11944 located at S12$^{\circ}$W08$^{\circ}$ and peaked at 18:32 UT.  Figure \ref{aia}a shows the {\it Solar Dynamics Observatory} \citep[\textit{SDO:}][]{pesnell2012}  {\it Atmospheric Imaging Assembly} \citep[AIA:][]{lemen2012}  193\,\AA\, difference image at 19:00 UT from which the 18:00 UT image is subtracted (prior to flare). The SDO/{\it Helioseismic and Magnetic Imager} \citep[HMI:][]{scherrer2012} colorized magnetogram image is superimposed on this difference image.  In the colorized magnetogram overlay, yellow-red colors indicate negative polarity, and green-blue colors indicate positive polarity. Bright post-eruption loops are seen near the active region with an extensive dimming region visible to the south and southwest, with weak negative polarity. Figure \ref{aia}b shows the SDO/AIA 211\AA~ image near 18:00 UT with open (pink: negative polarity, green: positive polarity) and closed (white) magnetic field lines from a Potential Field Source Surface (PFSS) model overlaid \citep{schrijver2006}. Two large coronal holes are visible to the northeast of the active region, and open coronal field is present to the west of the region.

\begin{figure*}
\epsscale{1.0}
\plotone{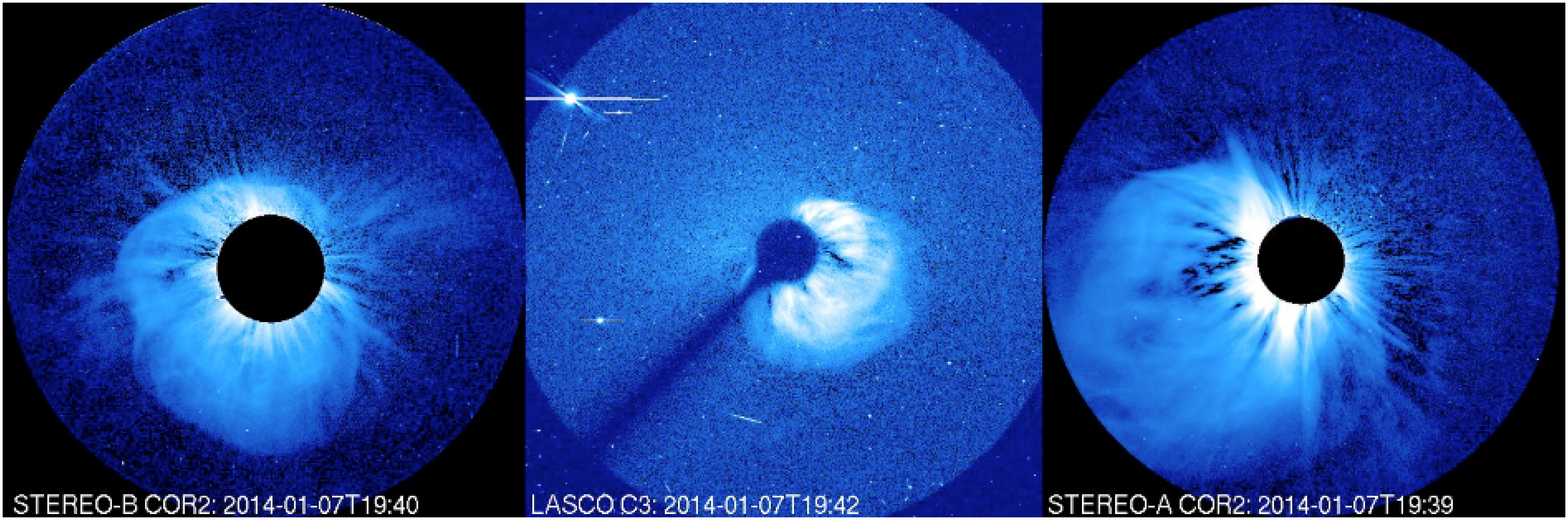}\plotone{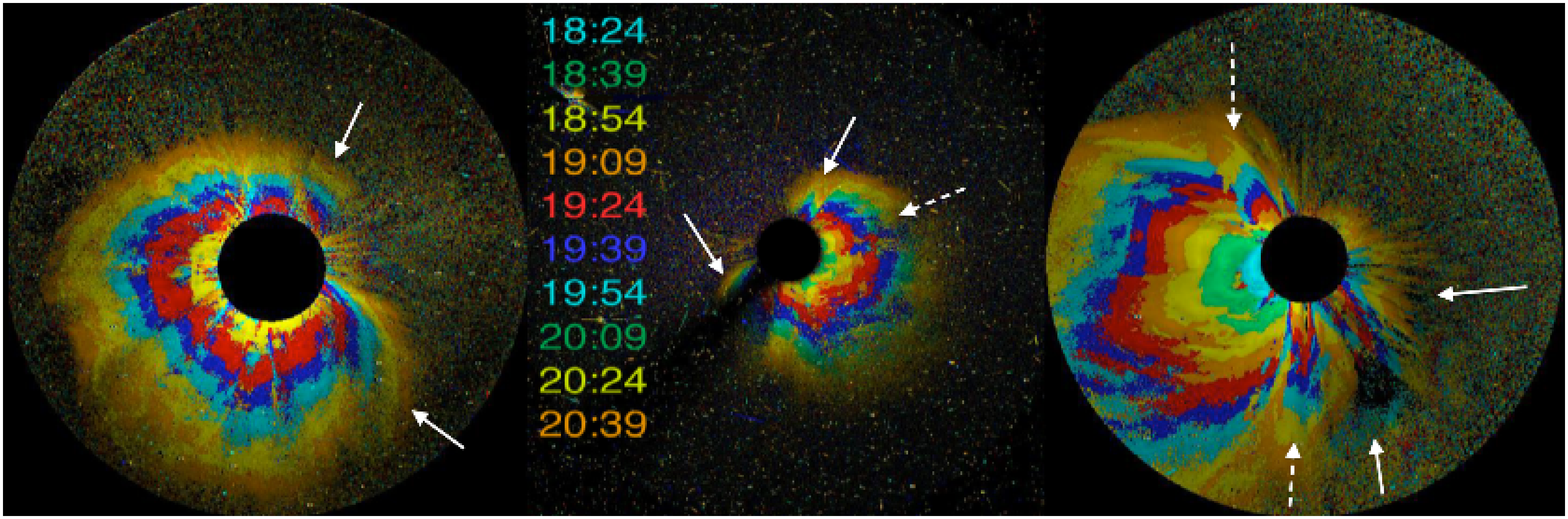}
\caption{Top row: Coronagraph observations of the 7 January 2014 CME as viewed from (a) {\it STEREO-B} SECCHI/COR2, (b) {\it SOHO} LASCO/C3, and (c) {\it STEREO-A} SECCHI/COR2, near the time of 19:40 UT. The fields-of-view of LASCO C3 and SECCHI/COR2 are 2.8-32 R$_{\odot}$ and 2.5-15 R$_{\odot}$ respectively. Bottom row: Observations from the same instruments with the Time Convolution Mapping Method applied from 18:24--20:39 UT.  Each color represents a different {\it STEREO-A} SECCHI/COR2 image time from 18:24--20:39 UT, with {\it STEREO-B} SECCHI/COR2 and {\it SOHO} LASCO/C3 images synchronized within 6 minutes of the {\it STEREO-A} time.  The solid arrows point out sheath/shock-like structures identified with TCMM, and dotted line arrows point out other parts of the CME that show different kinematic profiles from the ``driver'' or ``main-body'' such as along the CME legs. \label{cor}}
\end{figure*}

The top row of Figure \ref{cor} shows the CME first observed at 18:24 UT on 7 January 2014 by the {\it SOlar and Heliospheric Observatory} \citep[\textit{SOHO:}][]{soho} {\it Large Angle and Spectrometric Coronagraph Experiment} \citep[LASCO:][]{lasco} C3 and {\it STEREO}/SECCHI COR2-A and -B coronagraphs near the time of 19:40 UT.  {\it STEREO-A} was located at 151$^{\circ}$ longitude (west of the Sun-Earth line), and {\it STEREO B}  at $-153^{\circ}$ longitude (east of the Sun-Earth line), in  Heliocentric Earth Equatorial (HEEQ) coordinates\footnote{In this coordinate system the $Z$ axis is aligned with the solar north rotation pole and the $X$ axis pointing toward the intersection between the solar equator and the solar central meridian as seen from Earth \citep{hapgood1992,thompson2006}.  HEEQ coordinates are related to Stonyhurst heliographic coordinates, with directions south of the origin represented by negative HEEQ latitudes, and directions east by negative HEEQ longitudes.}. Earth was located at 0$^{\circ}$ longitude, $-3.8^{\circ}$ latitude; Venus at $-2.6^{\circ}$ longitude, $-1.8^{\circ}$ latitude; and Mars at 52$^{\circ}$ longitude, $-5.4^{\circ}$ latitude.

Three other energetic CMEs occurred in the days leading up to 7 January.  An Earth-directed CME originating from near active region number 11943 was first observed on 4 January 2014 at 19:39 UT by {\it STEREO-A}/SECCHI COR2 and was associated with an M4.0 flare from S09$^{\circ}$E34$^{\circ}$ peaking at 19:46 UT. The CME was propagating $\approx$\,4$^{\circ}$ west of the Sun-Earth line, $-35^{\circ}$ south of the solar equatorial plane, with a speed of $\approx$\,750\,$\pm$\,200 km\,s$^{-1}$. Later, on 7 January at 14:24 UT this CME arrived at Earth but did not produce a geomagnetic storm.  Shortly after, on 4 January at 23:08 UT ({\it STEREO-A}/SECCHI COR2) a CME erupted from active region 11936 just beyond the west limb with a speed of $\approx$\,650\,$\pm$\,100 km\,s$^{-1}$ associated with a partially occulted M1.9 flare.  A small increase in the background solar energetic particle flux was observed near Earth by the {\it Geostationary Operational Environment Satellites) (GOES)-13}/{\it  Electron, Proton, Alpha, and Detector} (EPEAD) starting on 5 January. Another CME originating from just beyond the west limb was observed on 6 January 2014 at 08:00 UT by {\it SOHO}/LASCO C2 from the same active region (11936), with an average speed of $\approx$\,1100 km\,s$^{-1}$.  The CME reached a peak speed of $\approx$\,1960 km\,s$^{-1}$ and was associated with a partially occulted C2.2 flare, a solar energetic particle event, and a ground level enhancement \citep{thakur2014}.  The solar energetic particle intensity remained enhanced before increasing again due to the X1.2 solar flare and 7 January CME, to above 1\,pfu/MeV (1 pfu\,=\,1 particle\,cm$^{-2}$\,sr$^{-1}$\,s$^{-1}$) as observed near Earth by the {\it GOES-13}/EPEAD 15\,-\,40\,MeV proton channel.

\subsection{Time Convolution Mapping Method}\label{tcmm}
 In this study, the CME ejecta (driver) is the input to the WSA--ENLIL+Cone model, therefore it is important to separate the CME ejecta from CME-associated brightenings such as streamer deflections and compressive wave fronts when interpreting the coronagraph images.  We applied a CME identification algorithm, Time Convolution Mapping Method (TCMM; \citet{thompson2015b,thompson2015a}), to the coronagraph data (top row of Figure \ref{cor}) to identify the time history of the CME evolution. The advantage to TCMM is that the spatiotemporal evolution of a CME can be captured, allowing users to separate features with different propagation characteristics. For example, separating ``true" or ``main" CME mass from CME-associated brightenings is a well-known obstacle \citep{gopalswamy2009,kwon2015}.  A TCMM CME map is made by first recording the maximum value each individual pixel in the image reaches during the traversal of the CME (in this case from 18:24 UT to 20:39 UT). Then the maximum value is convolved with a color to indicate the time that the pixel reached that value.   The lower row of Figure \ref{cor} shows the TCMM maps corresponding to the viewpoints in the top row for  multiple time intervals during the 7 January 2014 CME propagation. The TCMM user is then able to identify continuous ``rainbow profiles," indicating related kinematic behavior, and also identify breaks in the rainbow profiles that indicate a discontinuity in kinematic history.  Each color represents a different {\it STEREO-A} SECCHI/COR2 image time from 18:24--20:39 UT, with {\it STEREO-B} SECCHI/COR2 and {\it SOHO} LASCO/C3 images synchronized within 6 minutes of the {\it STEREO} time. Continuous loop-like TCMM structure (most visible in the lower right {\it STEREO-A} frame of Figure \ref{cor}) is indicative of the CME driver as described by \citet{gopalswamy2009}.   Kinematic rainbow profiles that appear as continuous loops are in contrast to the profiles with discontinuities that are more consistent with radial structures seen outside of the CME ejecta.  These discontinuities are more indicative of the interaction of the erupting structure with the surrounding corona.  This method also distinguishes the faint outer structure ahead of the bright CME loop, which some identify as the CME-driven shock front or compressive wave \citep{vourlidas2003,vourlidas2013,savani2013}.  \citet{kwon2015} have found evidence that many CMEs appear as full halos because of this surrounding structure, and not due to the viewing geometry of the main CME driver.  The solid arrows in Figure \ref{cor} point out sheath/shock-like structures identified with TCMM, and dotted line arrows point out other parts of the CME that show different kinematic profiles from the ``main body'', such as along the CME legs. The {\it STEREO-A} TCMM map also shows how the northern portion of the CME is deformed such that its motion is impeded compared with the southern CME front. 

\subsection{CME Kinematics}\label{kinematics}
 Many different techniques exist for deriving the CME parameters of location, speed, and size that are used as input to the WSA--ENLIL+Cone model.  In this section we consider a variety of techniques to create an input parameter distribution that adequately samples the error bar range for this CME.  Each distribution is then used as input to the ensemble simulations A and B which will be described in Section \ref{prop}.

\begin{figure*}
\epsscale{1.0}
\plotone{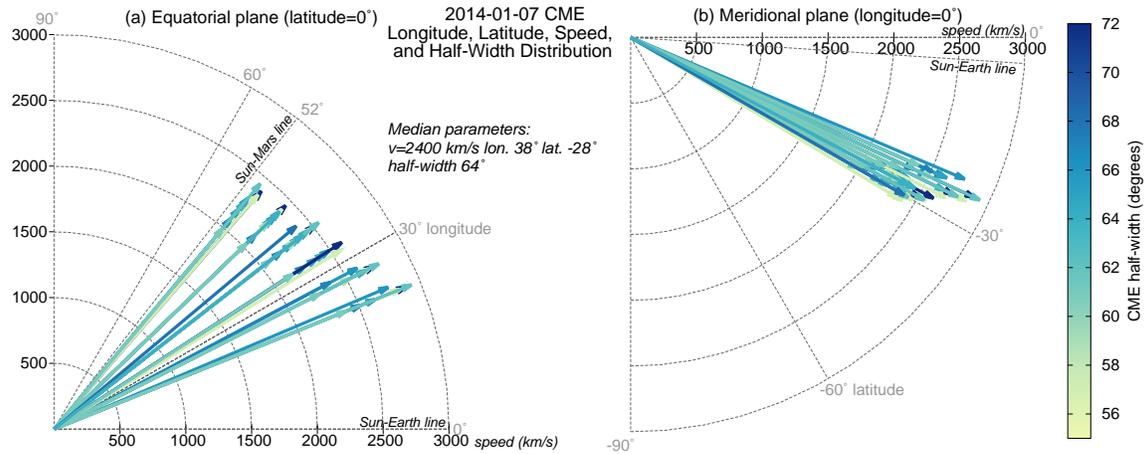}
\caption{Ensemble A: Distribution of the 7 January 2014 CME kinematic properties derived in real-time (using plane-of-sky estimates combined with the dimming location) shown in the (a) equatorial plane (latitude\,=0\,$^{\circ}$) and (b) meridional plane (longitude\,=\,0$^{\circ}$).  The plots show the CME speed vectors in spherical HEEQ coordinates with the grids showing the degrees longitude (a) and latitude (b), and the radial coordinate showing the speed in km\,s$^{-1}$. The Sun-Earth line is along 0$^{\circ}$ longitude and $-3.8^{\circ}$ latitude and the Sun-Mars line along 52$^{\circ}$ longitude and $-5.4^{\circ}$ latitude. Arrow directions indicate the CME central longitude and latitude, with the CME half-width indicated by the color of the vector, and arrow lengths correspond to CME speed. The CME propagation direction ranges from $-23^{\circ}$ to $-31^{\circ}$ latitude, 22$^{\circ}$ to 50$^{\circ}$ west of the Sun-Earth line in longitude, and speed ranges from $\approx$\,2000 to 3000 km\,s$^{-1}$.  Median CME parameters are: speed of 2400 km\,s$^{-1}$, direction of 38$^{\circ}$ longitude, $-28^{\circ}$ latitude, and a half-width of 64$^{\circ}$.\label{Apolar}}
\end{figure*}

CCMC derived a first round of true 3D parameters (de-projected) in real-time for the 7 January CME, using plane-of-sky measurements of beacon data from {\it SOHO}/LASCO-C3, {\it STEREO}/SECCHI COR2 combined with the CME direction that was estimated from the extensive dimming to the southwest of the active region.  The de-projected speed was calculated using the ratio of projected speed to the true speed relation from the theoretical geometric model by \citet{hundhausen1994} using the CME direction and projected CME half-width (57$^{\circ}$-72$^{\circ})$. With this technique the CME propagation direction was measured to range from $-23^{\circ}$ to $-31^{\circ}$ latitude, 22$^{\circ}$ to 50$^{\circ}$ west of the Sun-Earth line in longitude, and speed from $\approx$\,2000 to 3000 km\,s$^{-1}$.  Median CME parameters are: speed of 2400 km\,s$^{-1}$, direction of 38$^{\circ}$ longitude, $-28^{\circ}$ latitude, and a half-width of 64$^{\circ}$. These kinematic parameters are extrapolated for a height of 21.5 R$_{\odot}$, the inner boundary of the ENLIL model. Discrete samples from the range of measured CME parameters are shown in Figure \ref{Apolar} in the (a) equatorial plane (latitude\,=\,0$^{\circ}$) and (b) meridional plane (longitude\,=\,0$^{\circ}$)  and is used as input to ensemble A.  The plots show the CME velocity vectors in spherical HEEQ coordinates, the grids show degrees longitude (a) and latitude (b), and the radial coordinate shows the speed in km\,s$^{-1}$. The Sun-Earth line is along 0$^{\circ}$ longitude and $-3.8^{\circ}$ latitude. Arrow directions indicate the CME central longitude and latitude, with the CME half-width indicated by the color of the vector, and arrow lengths correspond to CME speed.

\begin{table*}
\tabletypesize{\scriptsize}
\caption{Summary of de-projected CME parameters from various techniques described in Section \ref{kinematics}.\label{tbl:meas}}
\begin{center}
\begin{tabular}{lcccccc}
\tableline
\tableline
Technique   & \multicolumn{1}{c}{$v$} & Lat & Long &  $\alpha/2$  & $\beta/2$ & Tilt \\
 & \multicolumn{1}{c}{{\scriptsize(km\,s$^{-1}$)}} & {\scriptsize($^{\circ}$)}   & {\scriptsize($^{\circ}$)}  & {\scriptsize($^{\circ}$)}  & {\scriptsize($^{\circ}$)}  & {\scriptsize($^{\circ}$)} \\
\tableline
\multicolumn{7}{l}{Real-time (Ensemble A):} \\
~~\citet{hundhausen1994} geometric model & 2400\,$^{+\,527}_{-\,410}$ & $-28\,^{+\,3}_{-\,5}$ & $38\,^{+\,12}_{-\,16}$ & $64\,\pm\,8$ & - & - \\
Science level data:& & & & & & \\
~~GCS & $2565\,\pm\,250$ & $-25\,\pm\,5$ & $32\,\pm\,10$ & $35\,\pm\,4$ & $22\,\pm\,4$ & $38\,\pm\,5$ \\
~~Fixed-$\phi$ and Harmonic Mean$^*$ (triangulation)& $2124\,\pm\,283$ & -- & $45\,\pm\,10$ & -- & -- & -- \\
~~Spherical Shell & $2000\,\pm\,200$ & $-28\,\pm\,2$ & $32\,\pm\,2$ & -- & -- & -- \\
\multicolumn{7}{l}{Combination of methods (Ensemble B):} \\
 & $2157\,^{+\,528}_{-\,340}$& $-25\,\pm\,5$ & $36\,\pm\,14$ & $44\,\pm\,12$ & $28\,\pm\,8$  & $38$\\
\tableline
$^*$parameters are for the ecliptic component of the CME.
\end{tabular}
\end{center}
\end{table*}

\begin{figure*}
\epsscale{1.0}
\plotone{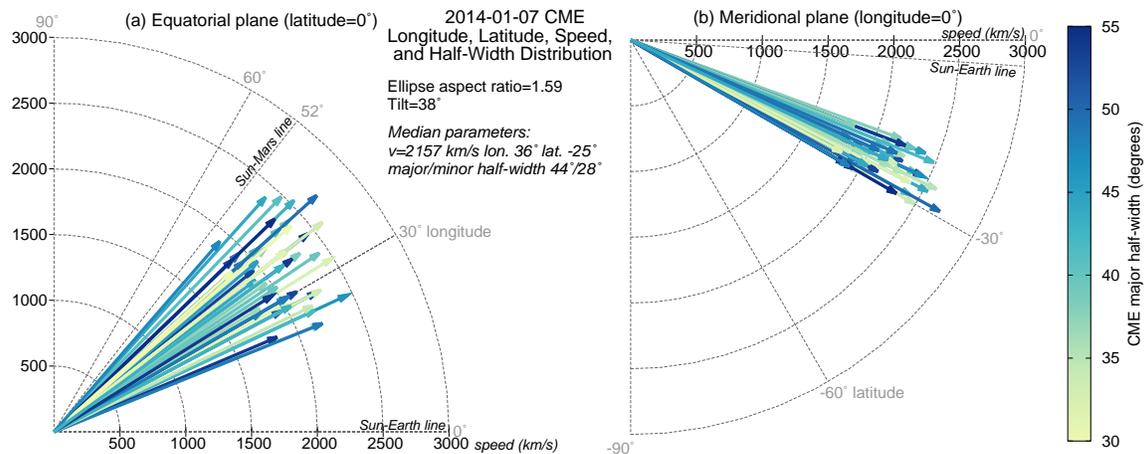}
\caption{Ensemble B:  Distribution of the 7 January 2014 CME kinematic properties derived from the GCS model and other analysis techniques, in the same format as Figure \protect\ref{Apolar}. The CME ellipse aspect ratio is 1.59 with the major axis half-width indicated by the color, and CME tilt is 38$^{\circ}$ counter clock-wise relative to the solar equator. The CME propagation direction ranges from $-20^{\circ}$ to $-30^{\circ}$ latitude, 22$^{\circ}$ to 49$^{\circ}$ west of the Sun-Earth line in longitude, and speed ranges from 1817 to 2685 km\,s$^{-1}$.   Median CME parameters are: speed of 2157 km\,s$^{-1}$, direction of 36$^{\circ}$ longitude, $-25^{\circ}$ latitude, major half-width of 44$^{\circ}$, and minor half-width of 28$^{\circ}$.\label{Bpolar}}
\end{figure*}

 \citet{mostl2015}  discuss the 3D kinematics of the 7 January CME using science level data with a variety of analysis methods.  The techniques and parameters are summarized in Table \ref{tbl:meas}. Measurements between 2.1 and 18.5 R$_{\odot}$ using the Graduated Cylindrical Shell (GCS) model from \citet{thernisien2006} yielded a speed of 2565\,$\pm$\,250 km\,s$^{-1}$, 32$^{\circ}\,\pm\,$10$^{\circ}$ longitude, $-25^{\circ}\pm$5$^{\circ}$ latitude, $\alpha/2=$ 35$^{\circ}$\,$\pm$\,4$^{\circ}$ major axis half-width,  $\beta/2=$22$^{\circ}$\,$\pm$\,4$^{\circ}$ minor axis half-width, with a tilt of 38$^{\circ}$\,$\pm$\,5$^{\circ}$ counter clock-wise relative to the solar equator. (The error bars are based on multiple GCS fits and are thus particular to this CME configuration). The GCS model is a forward modeling technique which assumes a magnetic flux rope topology \citep{thernisien2006,thernisien2009}.  The ratio of the CME face-on width (ellipse major axis) to the edge-on width (ellipse minor axis) gives an ellipse  aspect ratio of 1.59\,$\pm$\,0.2. \citet{mostl2015} used the {\it in-situ} arrival-times at Earth and Mars as constraints in the Ellipse Evolution model (ElEvo) model and found the ellipse aspect ratio to be in the range of 1.4\,$\pm$\,0.4 (with no tilt).  The average results of the fixed-$\phi$ and Harmonic Mean triangulation techniques \citep{liu2010b} between 20--30 R$_{\odot}$ give a speed of 2124\,$\pm$\,283 km\,s$^{-1}$, and longitude of 45$^{\circ}$\,$\pm$\,10$^{\circ}$ for the ecliptic component of the CME \citep{mostl2015}.  We also fit a spherical shell model to this CME, and found a speed of 2000\,$\pm$\,200 km\,s$^{-1}$, 32$^{\circ}$\,$\pm$\,2$^{\circ}$ longitude and $-28^{\circ}\,\pm \,$2$^{\circ}$ latitude.

Figure \ref{Bpolar} (same format as Figure \protect\ref{Apolar}) shows the distribution of the 7 January CME kinematic properties at 21.5 R$_{\odot}$ from the analysis techniques described above and is used as input to ensemble B. This distribution samples the error bar range of the CME parameter space derived predominantly from the GCS technique, but also from HI analysis techniques (see Table \ref{tbl:meas}). The CME ellipse aspect ratio is 1.59 with the major axis half-width indicated by the color, and CME tilt is 38$^{\circ}$ counter clock-wise relative to the solar equator.  The CME propagation direction ranges from $-20^{\circ}$ to $-30^{\circ}$ latitude, 22$^{\circ}$ to 49$^{\circ}$ west of the Sun-Earth line in longitude, and speed ranges from 1817 to 2685 km\,s$^{-1}$. The major and minor half-widths range from 22$^{\circ}$ to 54$^{\circ}$ and 13$^{\circ}$ to 32$^{\circ}$ respectively.  Median CME parameters are: speed of 2157 km\,s$^{-1}$, direction of 36$^{\circ}$ longitude, $-25^{\circ}$ latitude, major half-width of 44$^{\circ}$, and minor half-width of 28$^{\circ}$.

\section{Interplanetary Propagation}\label{prop}
To study the interplanetary propagation and arrival-time of this CME we use the ensemble WSA--ENLIL+Cone model from the CCMC. The global 3D MHD ENLIL model provides a time-dependent description of the background solar wind plasma and magnetic field using the WSA coronal model \citep{arge2000,arge2004} as input at the inner boundary of 21.5 R$_{\odot}$ \citep{ods1996,odstrcil1999_1,odstrcil1999_2,odstrcil2003,odstrcil2004}.   A homogeneous, over-pressured hydrodynamic plasma cloud is launched through the inner boundary of the heliospheric computational domain and into the background solar wind. The modeled CME cloud is approximated by a sphere, however ENLIL also supports other CME shapes such as an ellipsoid that can have an arbitrary tilt with respect to the solar equator.  These shapes can also be elongated in the radial direction, and different leading and trailing edge velocities are also possible. In this study, a spherical CME cloud (default) and tilted ellipsoid cloud are used. Cloud parameters of location, speed and size can be determined using any technique for deriving CME kinematic properties from coronagraph data.  A common technique is to assume that the geometrical CME properties are approximated by the Cone model \citep{zhao2002,xie2004} which assumes isotropic expansion, radial propagation, and constant CME cone angular width.

Ensemble modeling of a CME is a collection of CME arrival-time predictions that represent possible scenarios given the uncertainties associated with initial conditions from observations and modeling \citep{lee2013,emmons2013,mays2015}.  In addition to giving an average CME arrival-time prediction, this probabilistic forecast allows one to derive an arrival-time uncertainty from the spread and distribution of predictions.  The probability of CME arrival can be derived from the number of CME arrival predictions as a percentage of ensemble size.  The current ensemble modeling implementation at CCMC evaluates the sensitivity of WSA--ENLIL+Cone model simulations of CME propagation to initial CME parameters. For $n$ initial CME parameters, $n$ simulations are performed that provide $n$ profiles of MHD quantities (density, velocity, temperature, and magnetic field) and a distribution of $n$ model arrival-times at locations of interest within the computational domain.    For each simulation and location of interest, a CME arrival ``hit'' is defined by an increase in the dynamic pressure.  The ratio $n_{\rm hits}/n$ gives a forecast probability and conveys the forecast uncertainty about whether or not a CME will arrive at that location.  For this study $n$\,=\,48 ensemble members are used such that 48 individual simulations compose each ensemble.  Details of the WSA--ENLIL+Cone ensemble modeling system at the CCMC are described in \citet{mays2015}.

For Earth-directed CMEs, $n$ estimates of the geomagnetic $K_{\rm P}$ index are computed using the WSA--ENLIL+Cone model plasma parameters at Earth with a modified \citet{newell2007} coupling function \citep{savani2015b,mays2015}.  The function represents the rate of magnetic flux (d${\Phi_{\rm MP}}/$d${t}$) opening at the magnetopause d${\Phi_{\rm MP}}/$d${t} = {v_{\rm bulk}}^{4/3} B_{\rm T}^{2/3}{\sin}^{8/3}(\frac{\theta_{\rm C}}{2})$ where $v_{\rm bulk}$ is the bulk solar wind speed, the interplanetary magnetic field (IMF) clock angle $\theta_{\rm C}$ is given by $\tan^{-1}(B_y/B_z)$, and the perpendicular component of the magnetic field is given by $B_{\rm T}=(B_y^2+B_z^2)^{1/2}$ (in GSM coordinates).  An exponential fit to the correlation of this coupling function with the  $K{\rm p}$ index yields the relation $K_{\rm P} = 9.5 - e^{2.2-5.2 ({\rm d}{\Phi_{\rm MP}}/{\rm d}{t})}$. 

 We use the 7 January CME parameter distributions described in Section \ref{kinematics} as input to WSA--ENLIL+Cone ensemble simulations and compare these with observations in the following Sections \ref{ensa}-\ref{ensb}.   Ensemble A uses the default spherical CME cloud shape and the CME parameter distribution  derived in real-time using the \citet{hundhausen1994} geometric model (see Figure \ref{Apolar} and Table \ref{tbl:meas}). Ensemble B, in contrast uses an ellipsoid CME shape and an input parameter distribution predominantly based on the GCS technique (see Figure \ref{Bpolar} and Table \ref{tbl:meas}).  Both ensembles A and B use the same input WSA background solar wind map.

\begin{figure*}
\epsscale{.58}
\plotone{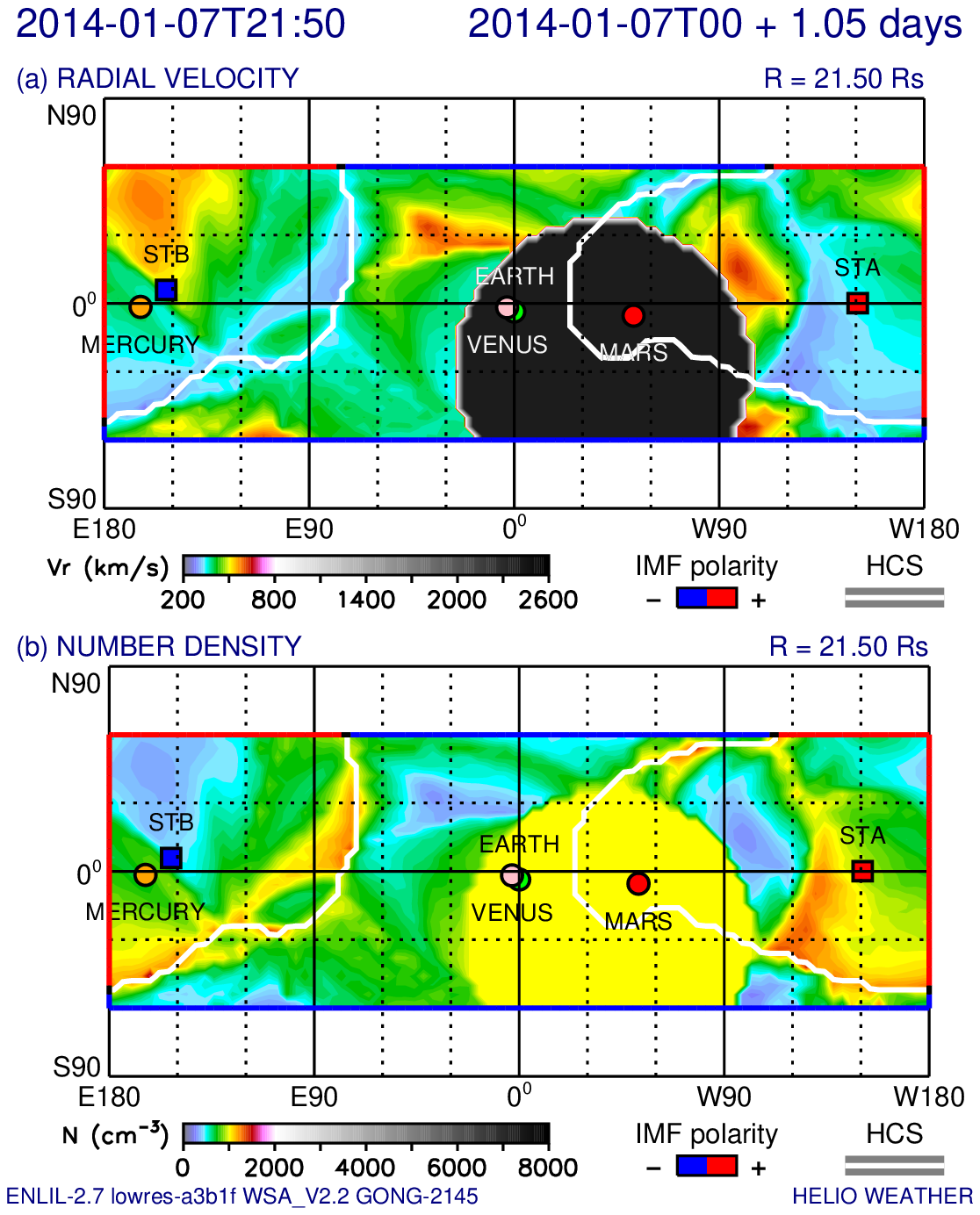}\plotone{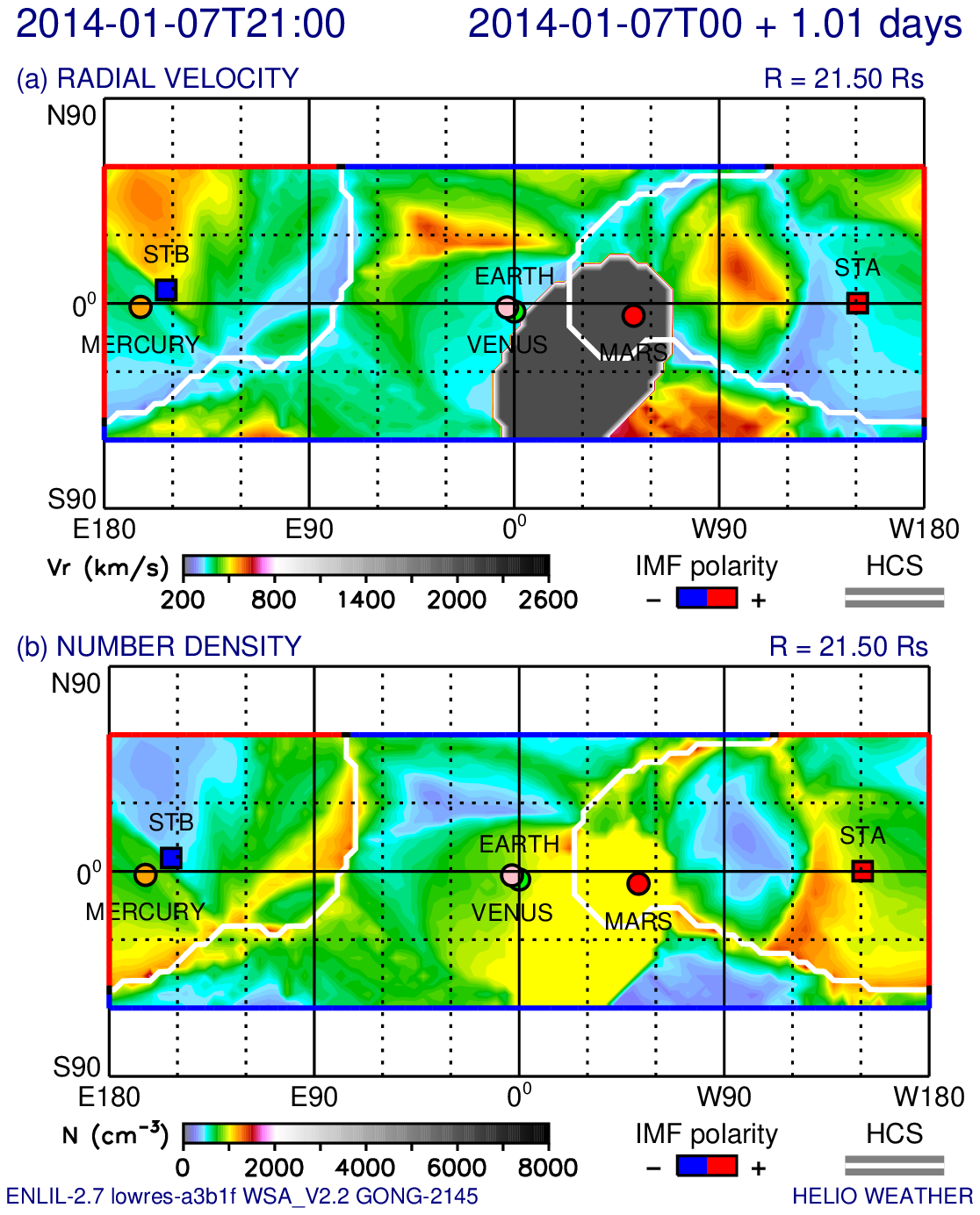}
\caption{Snapshot of the ENLIL model input inner boundary map at 21.5 R$_{\odot}$ computed from the WSA model for the (a) radial velocity and (b) number density used as input for one member of ensemble A (left) and ensemble B (right). The map shows the time when the inserted CME cross-section is largest, for the (left) spherical CME shape (equal to the input CME half-width) for ensemble A and (right) ellipsoid CME shape (equal to the input CME minor and major axis half-widths) from GCS measurements used in ensemble B.  Animations available in online version.\label{bnd}}
\end{figure*}

\subsection{Ensemble A: Real-time Measurements}\label{ensa}
\begin{figure*}  \centerline{\includegraphics[width=0.75\textwidth,angle=0,origin=c]{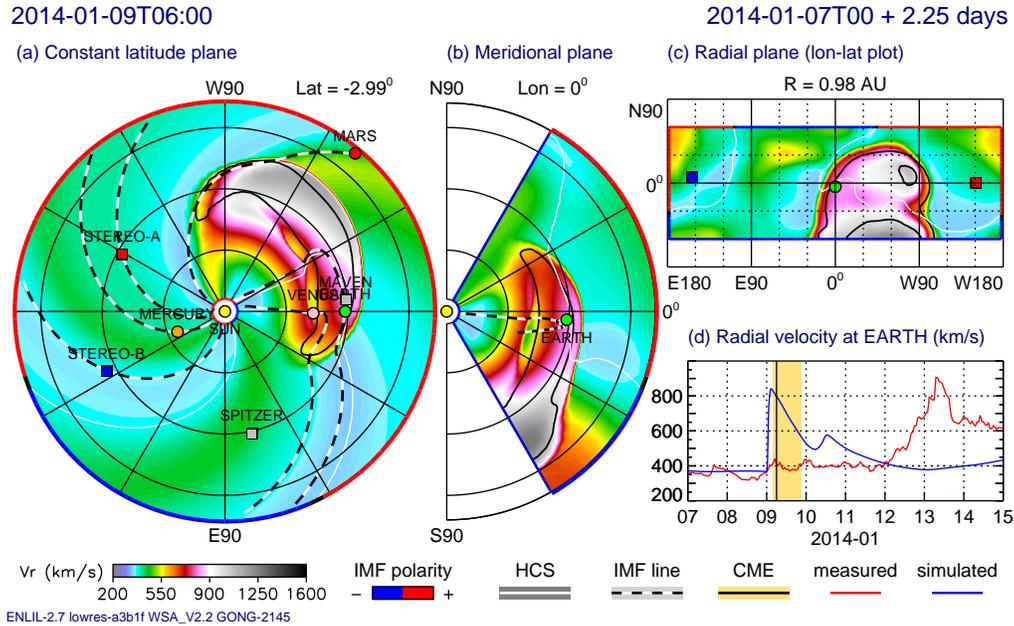}}
 \caption{Ensemble A global view of the 7 January 2014 CME on 9 January at 06:00 UT: WSA--ENLIL+Cone radial velocity contour plot for the (a) constant Earth latitude plane, (b) meridional plane of Earth, and (c) 1 AU sphere in cylindrical projection, for the ensemble member with median CME input parameters (speed of 2400 km\,s$^{-1}$, direction of 38$^{\circ}$ longitude, $-28^{\circ}$ latitude, and a half-width of 64$^{\circ}$). Panel (d) shows the measured (red) and simulated (blue) radial velocity profiles at Earth, with the simulated CME duration shown in yellow. Animation available in online version.\label{fig:timA}}
 \end{figure*}

CCMC completed the first ensemble simulation--A--for this CME in real-time (prior to CME arrival) using the input distribution shown in Figure \ref{Apolar} as initial conditions for 48 WSA--ENLIL+Cone simulations. A spherical CME shape was used, representing a symmetric CME angular width. Ensemble A uses the default settings available from CCMC's Runs on Request. The input WSA model map for this event was computed from a single National Solar Observatory {\it Global Oscillation Network Group} \citep[GONG:][]{harvey1996} daily-updated synoptic magnetogram for Carrington rotation number 2135 and Carrington longitude 40$^{\circ}$ on 7 January 2014 at 18:04 UT (Earth Carrington longitude was 100$^{\circ}$). Figure \ref{bnd} (left) shows a snapshot of the ENLIL model input inner boundary map at 21.5 R$_{\odot}$ computed from the WSA model for the (a) radial velocity and (b) number density in HEEQ coordinates, for ensemble A's median CME input parameters. The map shows the time when the inserted CME spheroid cross-section, a circle, is largest (equal to the input CME half-width) in the background solar wind. Red or blue outlines represent outward or inward IMF polarity respectively, and the white line shows the heliospheric current sheet (HCS).  Figure \ref{fig:timA} shows a radial velocity contour plot using this input for the (a) constant Earth HEEQ latitude plane, (b) meridional plane of Earth, and (c) 1 AU sphere in cylindrical projection on 9 January at 06:00 UT.   Panel (d) shows the measured OMNI hourly average (red) and simulated (blue) radial velocity profiles at Earth, with the simulated CME duration shown in yellow. The figure shows the northeastern portion of the CME impacting Earth. An animation of the input inner boundary map, and resulting simulation are available as electronic supplementary material.


\begin{figure*}
\epsscale{.58}
\plotone{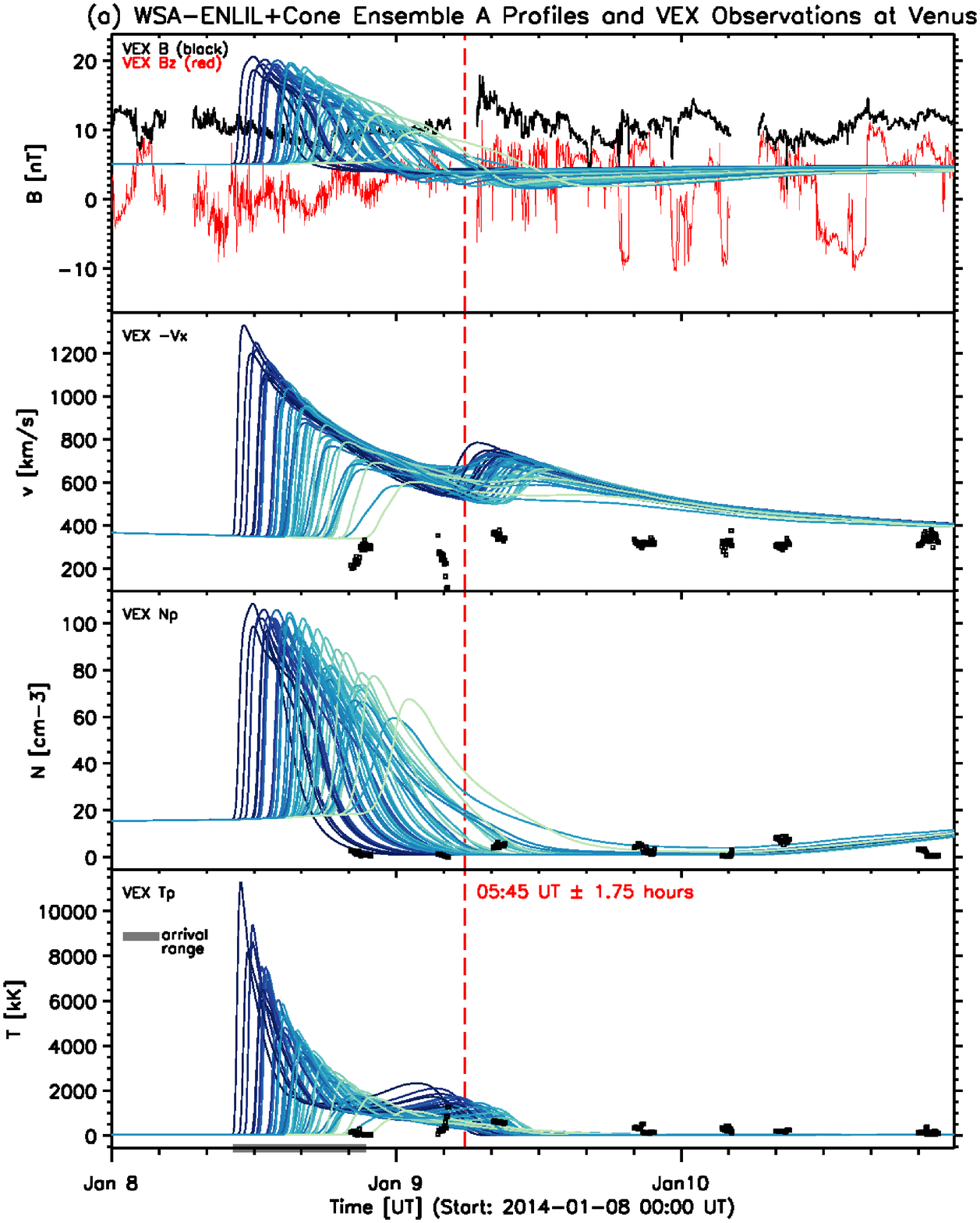}\plotone{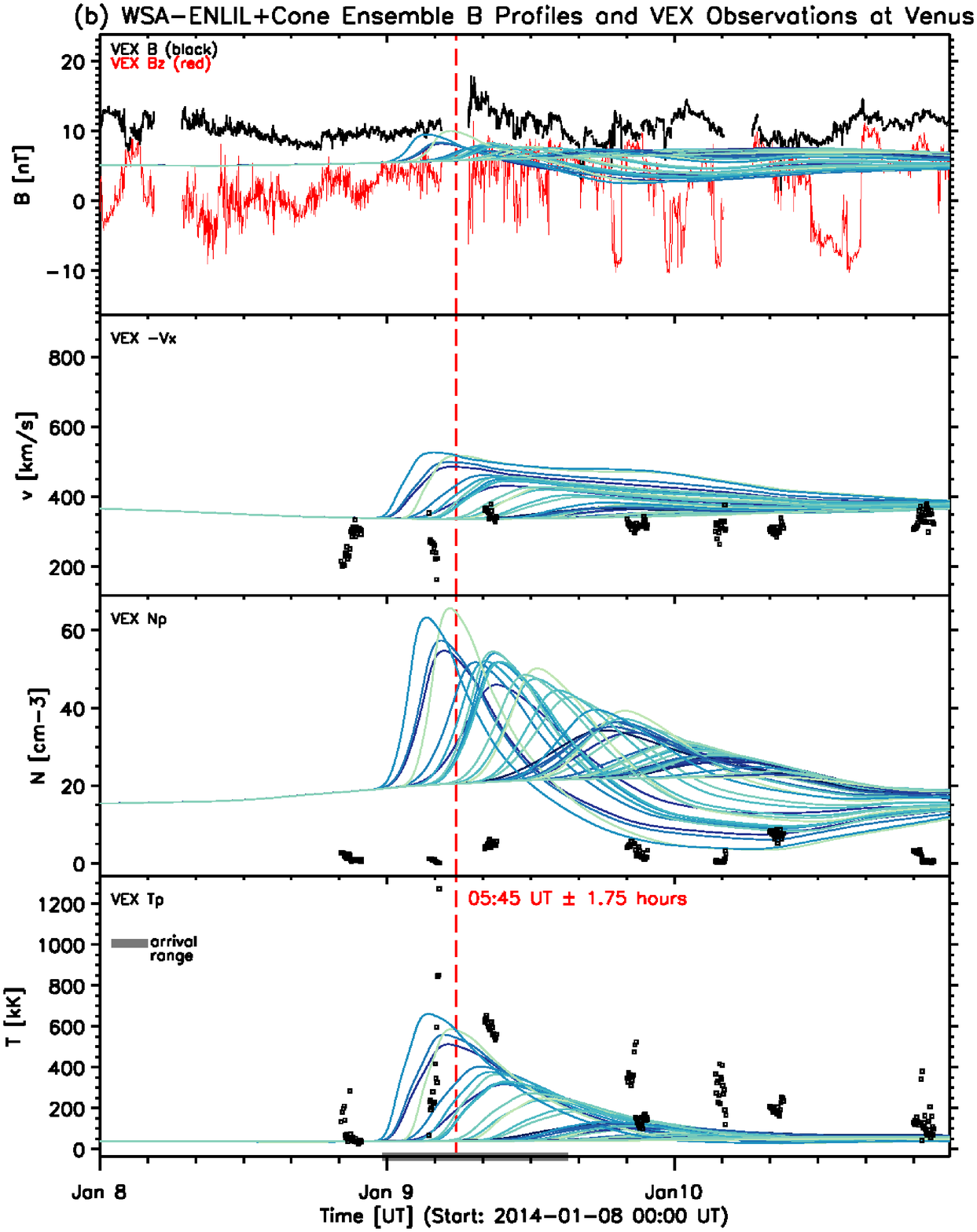}
\caption{\label{stacksVenus}Modeled magnetic field, radial velocity, density, and temperature profiles at Venus for (a) ensemble A and (b) ensemble B plotted as color traces (slow to fast CME input speeds range from light green to dark blue) for all 48 ensemble members.  {\it In-situ} observations from {\it VEX} are plotted black for periods when the spacecraft was in the solar wind. The total magnetic field is shown in black and $B_z$ is in red. The vertical dashed red line indicates that the beginning of the CME-associated disturbance is between 4:30--7:01 UT (5:45 UT $\pm$ 1.25 hours) on 9 January. The range of modeled arrival-times for hits is indicated by the grey bar on the time axis.}
\end{figure*}

\begin{figure*}
\epsscale{.58}
\plotone{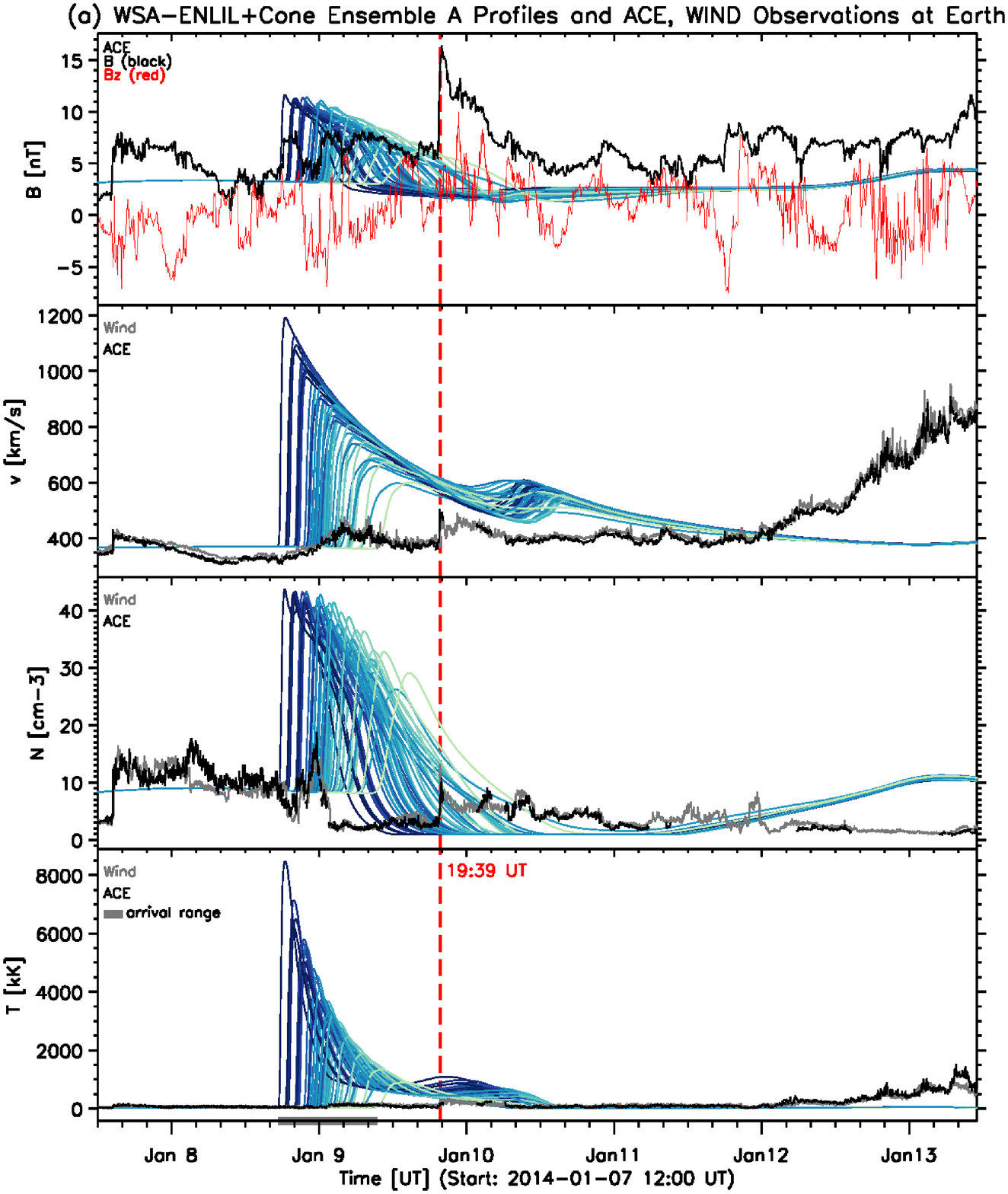}\plotone{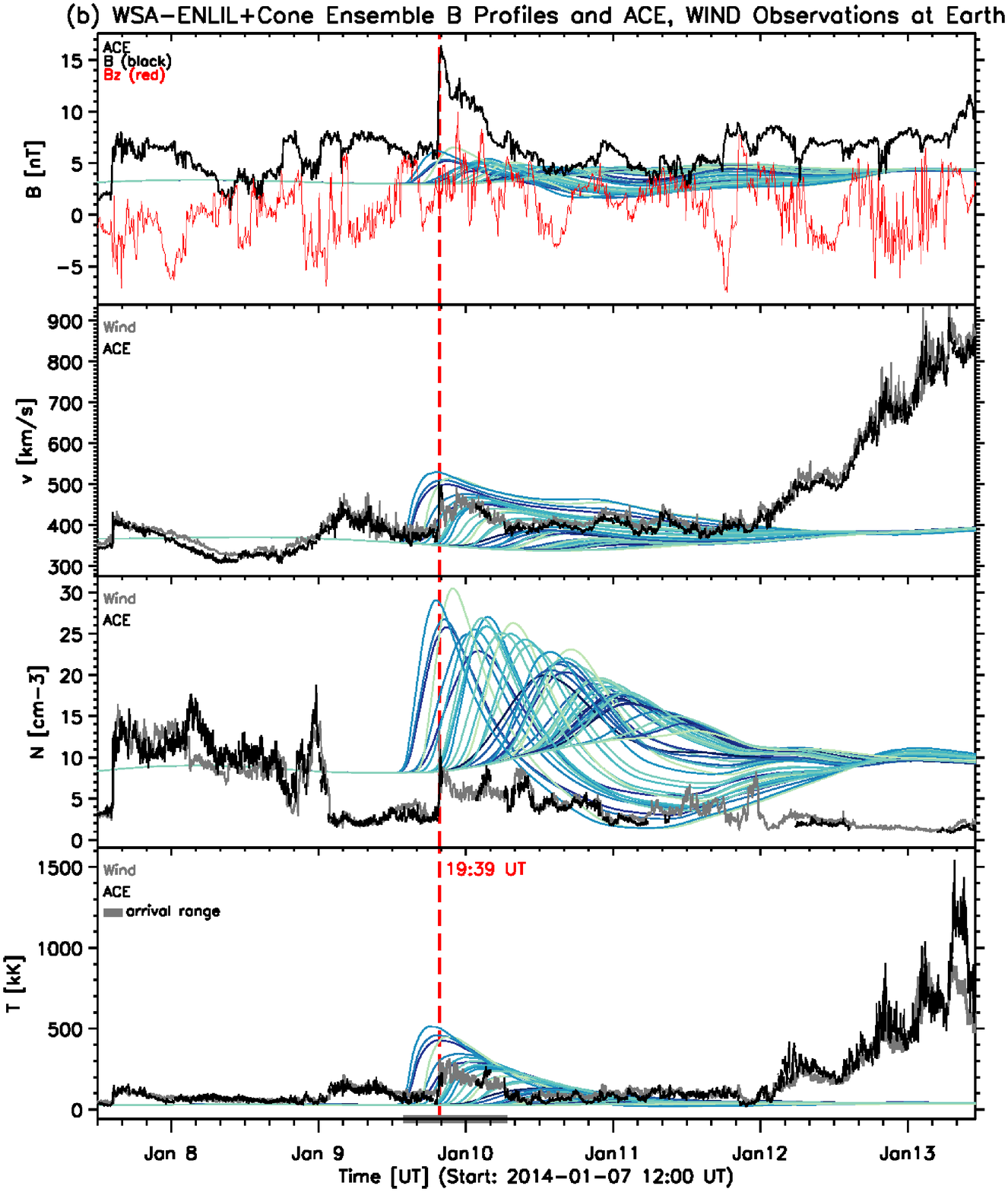}
\caption{\label{stacksEarth}Modeled magnetic field, radial velocity, density, and temperature profiles at Earth for (a) ensemble A and (b) ensemble B plotted as color traces (color coded by CME input speed) for all 48 ensemble members.  {\it In-situ} $L1$ observations from {\it ACE} are plotted in black (red for $B_z$) and {\it Wind} in grey.  The vertical dashed red line indicates the CME arrival observed on 9 January at 19:39 UT as a weak interplanetary shock. The range of modeled arrival-times for hits is indicated by the grey bar on the time axis.}
\end{figure*}

\begin{figure*}
\epsscale{.58}
\plotone{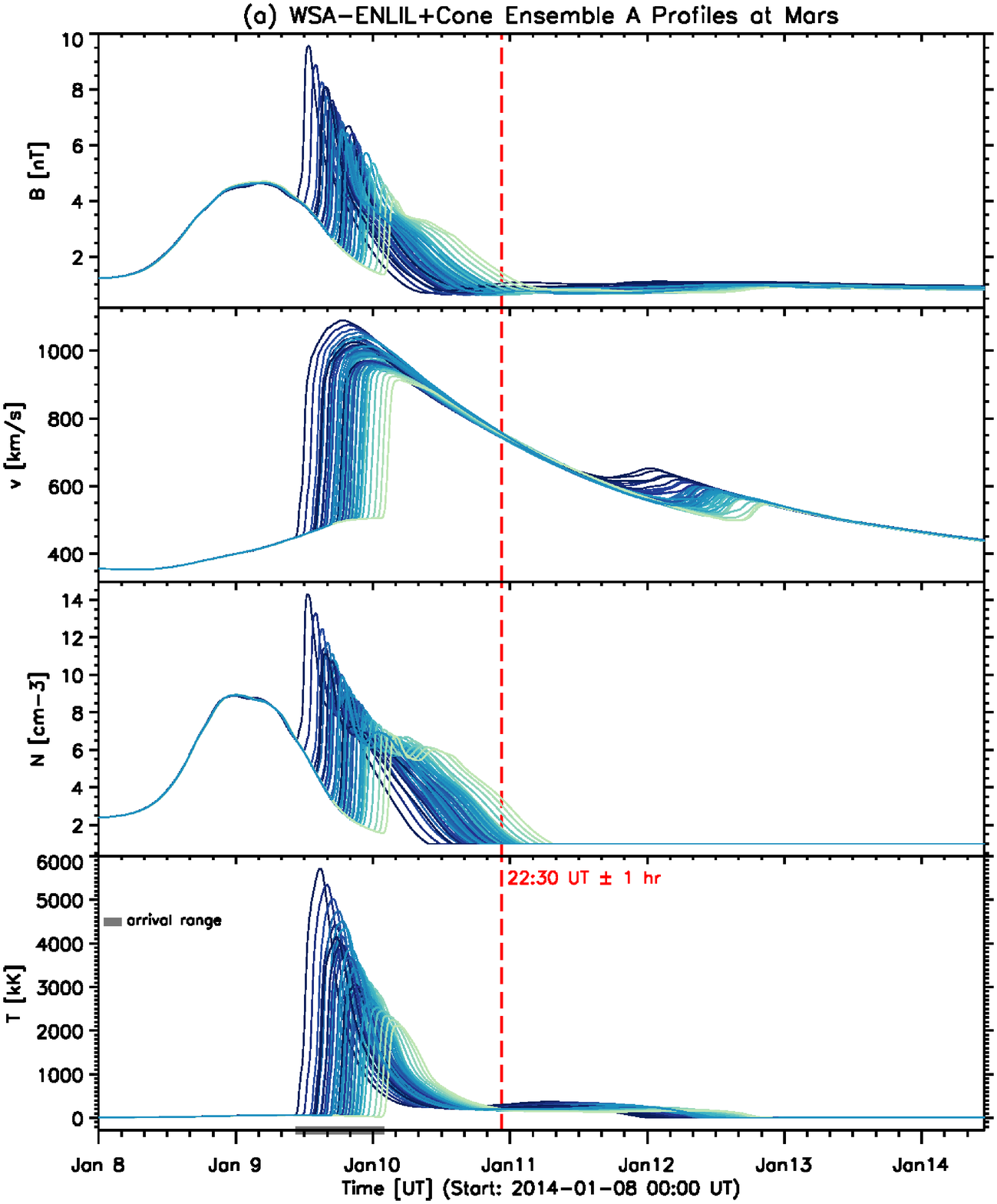}\plotone{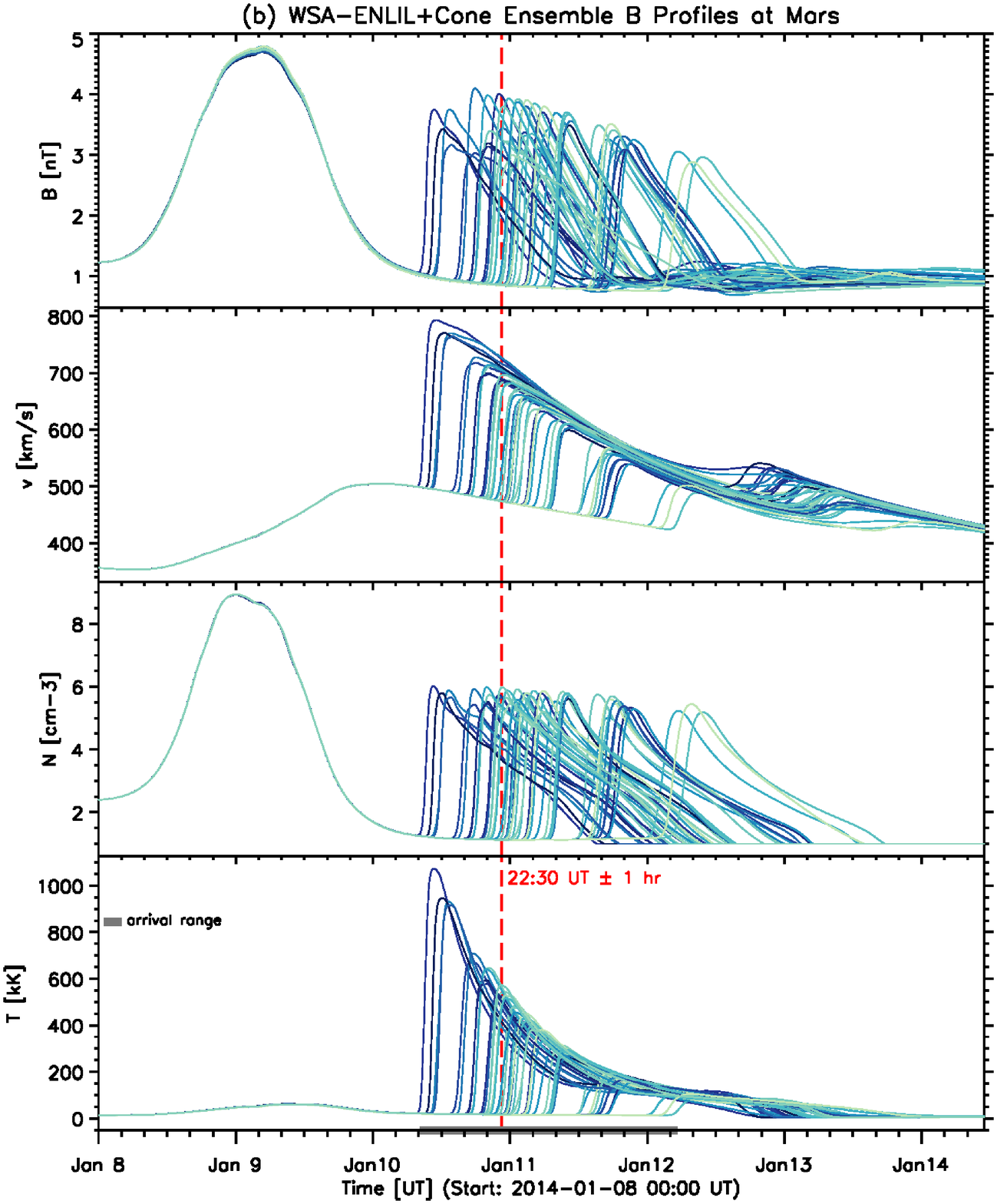}
\caption{\label{stacksMars}Modeled magnetic field, radial velocity, density, and temperature profiles at Mars for (a) ensemble A and (b) ensemble B plotted as color traces (color coded by CME input speed) for all 48 ensemble members.  The vertical dashed red line indicates the observed Forbush decrease time. The range of modeled arrival-times for hits is indicated by the grey bar on the time axis.}
\end{figure*}

Figures \ref{stacksVenus}a,  \ref{stacksEarth}a, and \ref{stacksMars}a show the modeled magnetic field, radial velocity, density, and temperature profiles at Venus, Earth, and Mars plotted as color traces for all 48 ensemble members.  The model traces are color-coded by CME input speed; slow to faster speeds range from light green to dark blue.  For Earth the {\it in-situ} L$_1$ observations from {\it Advanced Composition Explorer} \citep[\textit{ACE:}][]{ace} are plotted in black (red for $B_z$) and {\it Wind} is plotted in grey. At Venus, the {\it in-situ} observations from {\it Venus Express} \citep[\textit{VEX:}][]{svedhem2009} are shown for periods when the spacecraft was in the solar wind. For {\it VEX}, the total magnetic field is shown in black and $B_z$ in red. The range of model arrival-times for hits (defined by an increase in the model dynamic pressure) is indicated by the grey bar on the time axis of each figure. The results show that the  probability that the CME will arrive at Venus, Earth, and Mars is 100\% (48 hits out of 48 members).

Figures \ref{stacksVenus}a-\ref{stacksMars}a show that all ensemble A members arrive prior to the observed time at the three locations, indicated by the vertical dashed red line. First, {\it VEX} observed the beginning of the CME-associated disturbance between 4:30--7:01 UT (5:45 UT $\pm$ 1.25 hours) on 9 January, as shown in Figure \ref{stacksVenus}a.  The {\it VEX} data between about 9 January 4:32 UT (data gap follows) and 6:35 UT is during a Venus close encounter. After this period, a shock is observed with a prompt rise in magnetic field to about 19 nT at 7:01 UT. There is no plasma data available at, and about 4 hours around the shock.  After a sheath region, there are some magnetic field rotations observed from approximately 9 January 19:00 UT to 10 January 10:00 UT. However, the rotations are on a small angular scale and there is no well-defined flux rope or magnetic cloud observed. At Venus, the ENLIL ensemble mean arrival-time was on 8 January at 14:52 UT with individual arrival-times ranging from $+7/-5$ hours (indicated by the grey bar in Figure \ref{stacksVenus}a) around the mean. Using the mean arrival-time prediction, the prediction error is $\Delta t_{\rm err}=t_{\rm predicted}-t_{\rm observed}$\,=\,$-$14.9 hours (early prediction).   Figure \ref{stacksVenus}a shows that the modeled radial speed is a factor of $\approx$\,1.5 to 2.3 times higher than observed by {\it VEX}.

Next, the arrival of the CME was observed by {\it Wind} and {\it ACE} at $L1$ ahead  Earth on 9 January at around 19:39 UT as a weak interplanetary shock (see Figure \ref{stacksEarth}a).  The sheath region after the shock lasts until about 10 January 07:00 UT, during which the alpha particle abundance is increased. The magnetic obstacle is observed from approximately 10 January 07:00 UT to 11 January 03:00 UT, indicated by magnetic field rotations and bidirectional suprathermal electron flux (not shown). The magnetic field rotation directions indicated by the changes of cone and clock angles are roughly consistent with those observed at {\it VEX} (not shown). The rotations are on a larger scale at Earth than at {\it VEX}, however they are not smooth enough to be a magnetic cloud (lower-than-expected $\beta$ and proton temperature are also not observed). The mean ENLIL arrival-time at Earth was on 9 January at 00:10 UT with individual arrival-times ranging from $+9/-7$ hours (indicated by the grey bar in Figure \ref{stacksEarth}a) around the mean.  Using the mean arrival-time prediction, the prediction error is $-19.4$ hours (early prediction).  Figure \ref{stacksEarth}a shows that the modeled peak solar wind speed at the CME arrival-time ranged from 600 to 1200 km\,s$^{-1}$ while the observed value was $\approx$\,500 km\,s$^{-1}$. The density was also overpredicted by about a factor of 3 to 5.  The model also does not capture the high speed stream on 12 January reaching $\approx$\,850 km\,s$^{-1}$ from the coronal holes to the northeast of the active region (see Figure \ref{aia}b). The WSA--ENLIL background solar wind model driven by GONG synoptic magnetogram shows a slower $\approx$\,450 km\,s$^{-1}$ southern extension of the coronal hole crossing Earth on 14 January.  Figure \ref{stacksEarth}a also shows the observed arrival of an earlier CME from 4 January (described in Section \ref{obs}) at around 14:24 UT on 7 January.

\begin{figure*}
\epsscale{0.45}
\plotone{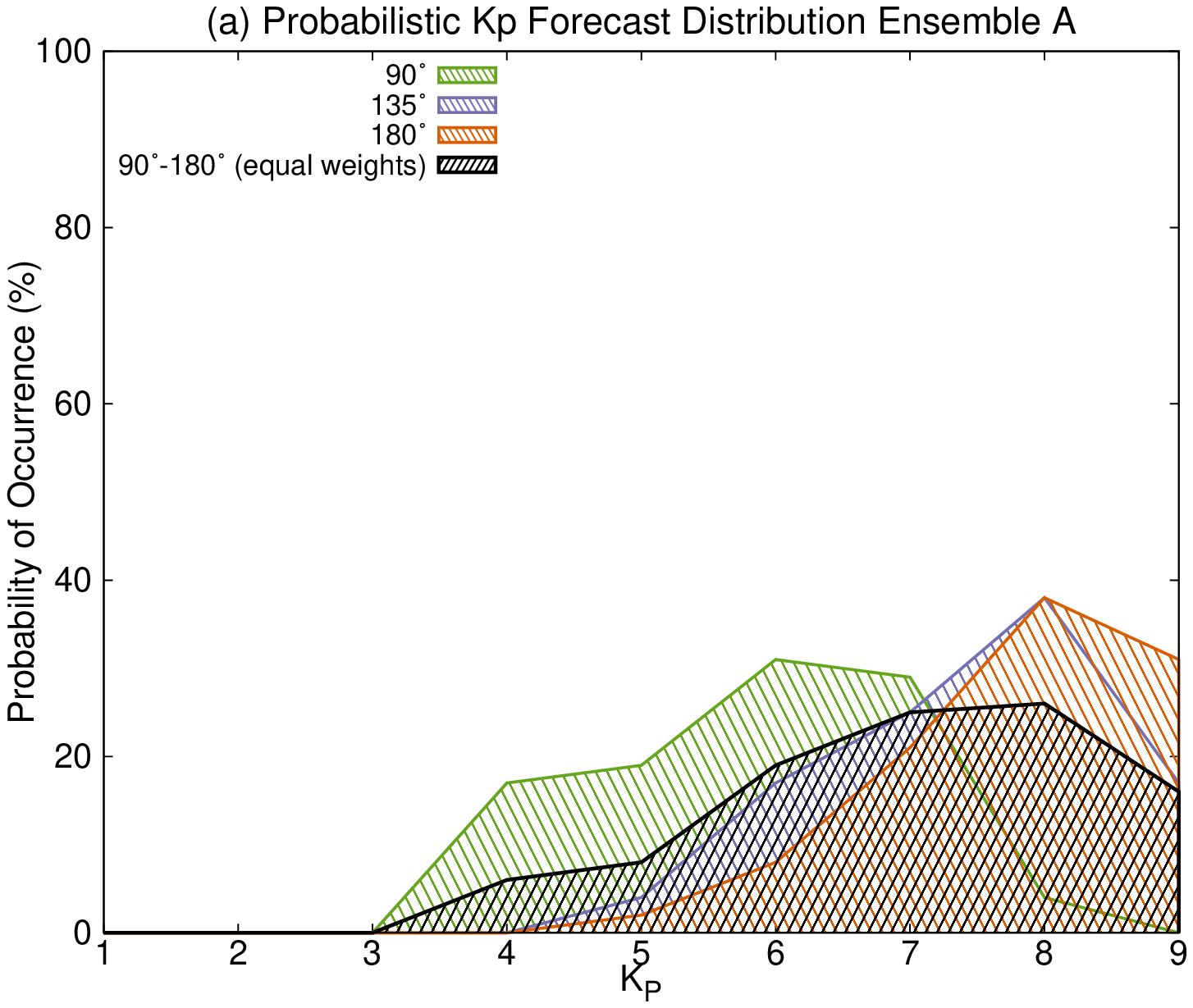}
\plotone{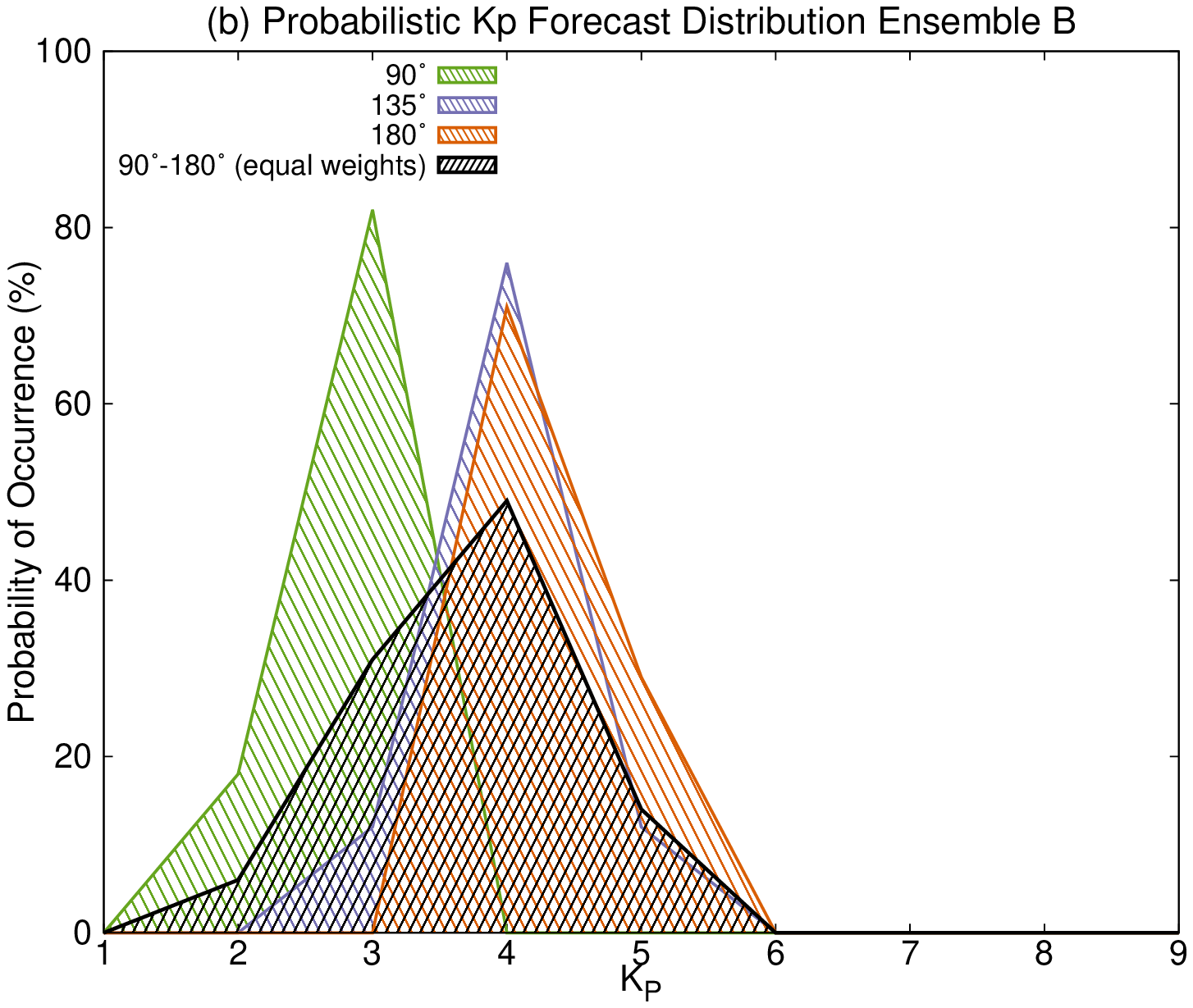}\\
\plotone{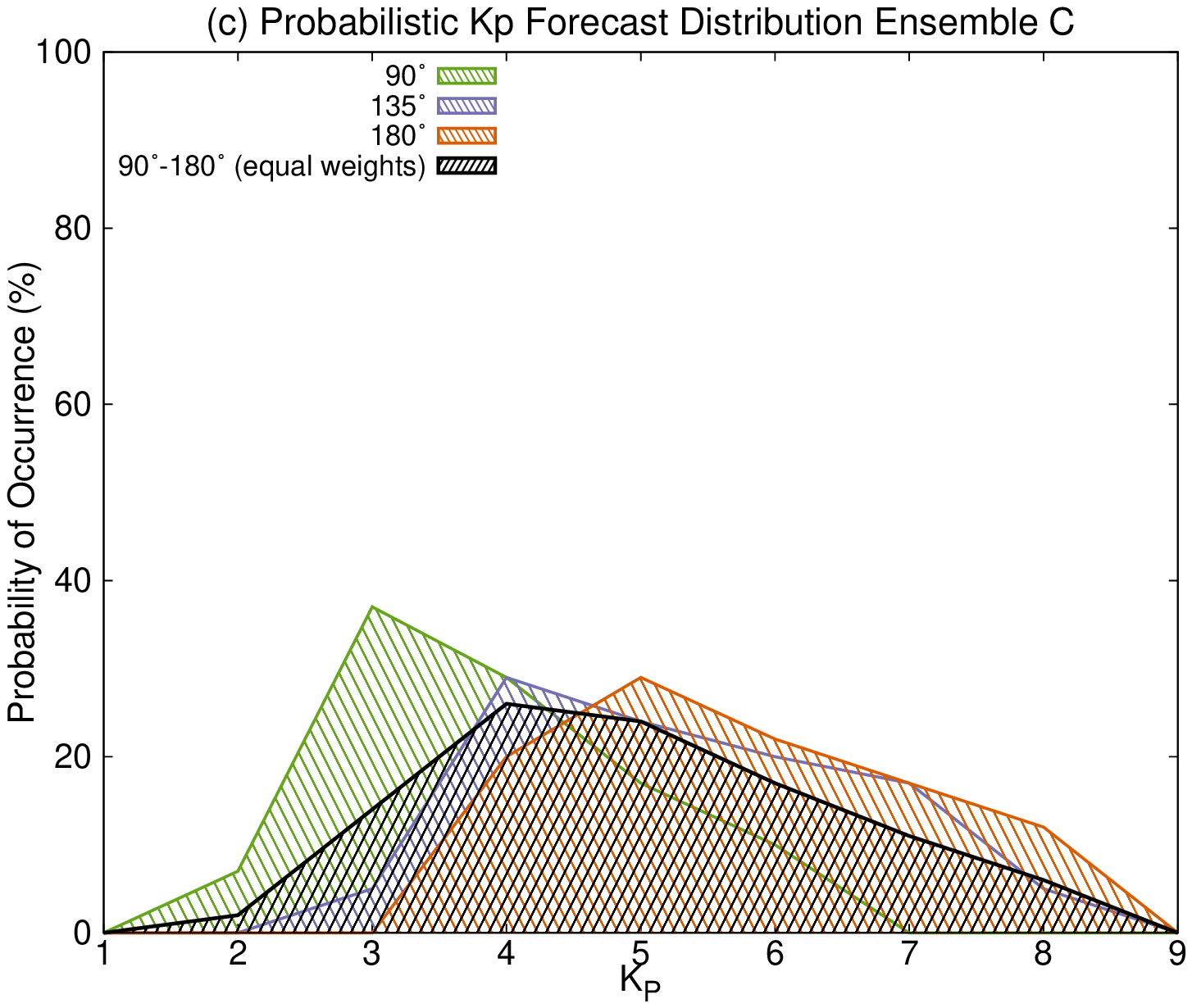}
\plotone{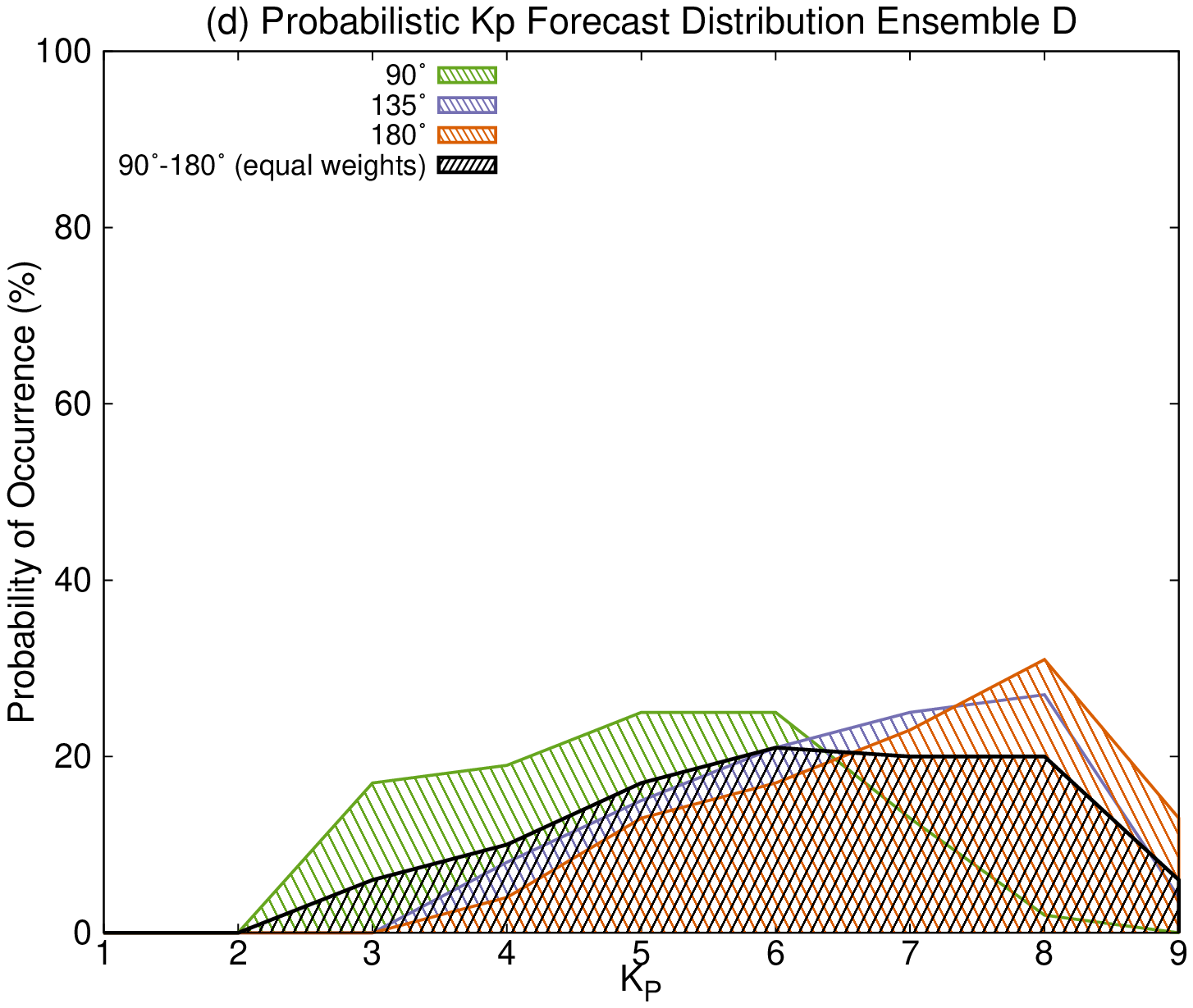}
\caption{\label{kpAB}Probabilistic $K_{\rm P}$ forecast distribution for (a) ensemble A, (b) ensemble B, (c) ensemble C, (d) ensemble D using ENLIL modeled solar-wind quantities at Earth for three IMF clock-angle scenarios $\theta_{\rm C}=90^{\circ}$ (green), 135$^{\circ}$ (purple), 180$^{\circ}$ (orange), and all three angles combined 90$^{\circ}$\,-\,180$^{\circ}$ (black) (assuming equal likelihood).  Ensemble A shows a 78\% chance for $K_{\rm P}$ to reach between 6--8 ($\sigma$=1.8), ensemble B a 94\% chance for $K_{\rm P}$ to reach between 3--5 ($\sigma$\,=\,0.8), ensemble C an 81\% chance for $K_{\rm P}$\,=\,3--7 ($\sigma$\,=\,1.5) and ensemble D (88\% chance for $K_{\rm P}$\,=\,4--8 ($\sigma$\,=\,1.6).}
\end{figure*}

The maximum $K_{\rm P}$ index was calculated using the ENLIL modeled quantities of the arrival at Earth (colored traces shown in \ref{stacksEarth}a).  Figure \ref{kpAB}a shows the predicted $K_{\rm P}$ probability distribution for three IMF clock angle scenarios $\theta_{\rm C}=90^{\circ}$ (green), 135$^{\circ}$ (purple), 180$^{\circ}$ (orange).  The figure also shows the overall predicted $K_{\rm P}$ probability distribution calculated for all three angles combined 90$^{\circ}$--180$^{\circ}$, assuming each equal weights, in black. The predicted $K_{\rm P}$ probability distribution has an overall standard deviation $\sigma$ of 1.8 and there is a $\approx$\,78\% chance for the maximum $K_{\rm P}$ to be between 6 and 8. Using the average $K_{\rm P}$ prediction of 7 or the most likely $K_{\rm P}$ prediction of 8, the $K_{\rm P}$ prediction error for this event is $\Delta K_{\rm P~err}= K_{\rm P~predicted}- K_{\rm P~observed}$\,=\,4.33 or 5.33 (overprediction).  For this ensemble simulation, the overprediction of the CME speed and density at Earth leads to an overprediction of $K_{\rm P}$.  \citet{mays2015} also discuss a general bias for the overprediction of $K_{\rm P}$ with this method for CME input speeds above $\approx$\,1000 km\,s$^{-1}$.

Next, a Forbush decrease was observed at Mars by the {\it Mars Science Laboratory} (MSL) Radiation experiment (RAD) at around 22:30 UT $\pm$ 1 hour on 10 January and the CME arrival is thought to be within 2 hours of the Forbush decrease time (as detailed by \citet{mostl2015}).  At Mars, the ENLIL ensemble mean arrival-time was on 9 January at 18:22 UT with individual arrival-times ranging from $+8/-8$ hours around the mean. Using the ensemble mean arrival-time, the prediction error is $-28.1$ hours (early prediction).  Figures  \ref{fig:timA} and \ref{stacksMars}a also shows a high speed stream modeled to arrive mid-day on 8 January, just ahead of the modeled CME arrival. Ensemble A modeled arrival-times are summarized in Table \ref{tbl:summ}.  

\subsection{Ensemble B: GCS Measurements}\label{ensb}

\begin{figure*}  
\centerline{\includegraphics[width=0.75\textwidth,angle=0,origin=c]{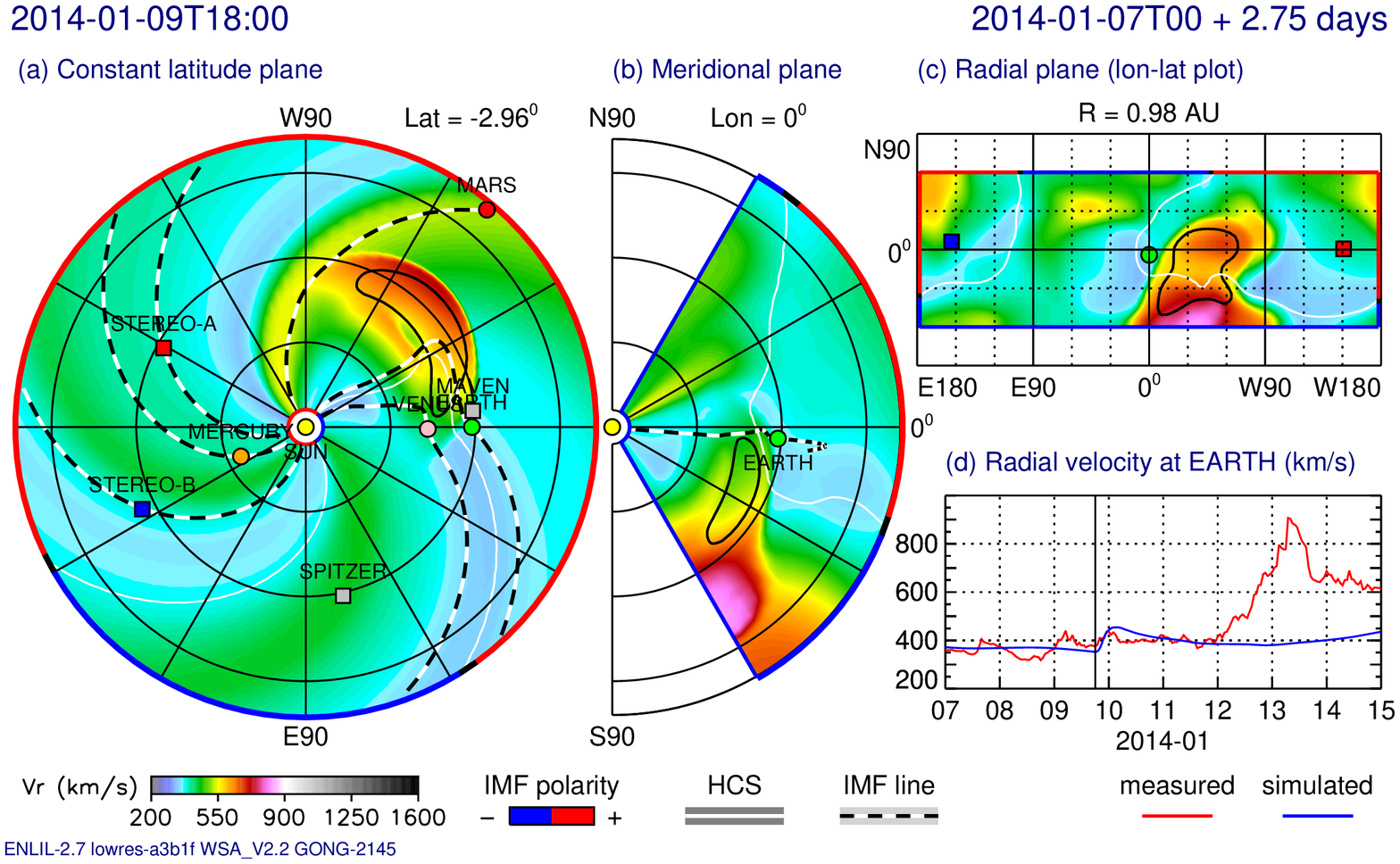}}
\caption{Ensemble B global view of the 7 January 2014 CME on 9 January at 18:00 UT: WSA--ENLIL+Cone radial velocity contour plot in the same format as Figure \ref{fig:timA} for one ensemble B member predicting a ``hit'' (CME input parameters: speed of 2061 km\,s$^{-1}$, direction of 30$^{\circ}$ longitude, $-24^{\circ}$ latitude, 49$^{\circ}$ major half-width, and 31$^{\circ}$ minor half-width). Panel (d) shows the measured (red) and simulated (blue) radial velocity profiles at Earth. Movie available in online version.\label{fig:timB}}
\end{figure*}  

\begin{table*}
\tabletypesize{\scriptsize}
\caption{Summary of ensemble modeling results for the 7 January 2014 CME at Venus, Earth, and Mars. Column 1: ensemble label. Columns 2-7: median CME parameters of speed $v$, latitude, longitude (HEEQ), major axis half-width $\alpha/2$, minor axis half-width $\beta/2$, and tilt.  The remaining columns list the ensemble model mean and spread in arrival-time and the prediction error $\Delta t_{\rm err}=t_{\rm predicted}-t_{\rm observed}$ for Venus, Earth, and Mars.  Ensemble A measurements and simulations were performed in real-time with a spherical CME shape. Ensemble B measurements and simulations were obtained with science level data and use an ellipsoid CME shape with a tilt from GCS measurements. Ensemble B' is identical to ensemble B except a spherical shape is used. Ensemble C half-widths are reduced by 20$^{\circ}$ compared with ensemble A. Ensemble D longitudes are increased by 10$^{\circ}$ compared with ensemble A.\label{tbl:summ}}
\begin{center}
\begin{tabular}{@{\extracolsep{3pt}}lrrrrrrlrlrlr@{}}
\tableline
\tableline
& & & & & & & \multicolumn{2}{c}{Venus} & \multicolumn{2}{c}{Earth} & \multicolumn{2}{c}{Mars}\\
\cline{8-9}\cline{10-11}\cline{12-13}
&\multicolumn{6}{c}{Median CME parameters} &  \multicolumn{1}{c}{Mean Arrival}  &  $\Delta t_{\rm err}$ &  \multicolumn{1}{c}{Mean Arrival}  &  $\Delta t_{\rm err}$ &  \multicolumn{1}{c}{Mean Arrival}  &  $\Delta t_{\rm err}$ \\
{\scriptsize Label} & \multicolumn{1}{c}{$v$} & Lat & Long &  $\alpha/2$  & $\beta/2$ & Tilt &  \multicolumn{1}{c}{Time~~~~Spread} &  & \multicolumn{1}{c}{Time~~~~Spread} & & \multicolumn{1}{c}{Time~~~~Spread} &\\
 & \multicolumn{1}{c}{{\scriptsize(km\,s$^{-1}$)}} & {\scriptsize($^{\circ}$)}   & {\scriptsize($^{\circ}$)}  & {\scriptsize($^{\circ}$)}  & {\scriptsize($^{\circ}$)}  & {\scriptsize($^{\circ}$)} & \multicolumn{1}{c}{{\scriptsize(UT)~~~~~~\scriptsize(h)}} & {\scriptsize(h)}  & \multicolumn{1}{c}{{\scriptsize(UT)~~~~~~\scriptsize(h)}} & {\scriptsize(h)} & \multicolumn{1}{c}{{\scriptsize(UT)~~~~~~\scriptsize(h)}} & {\scriptsize(h)}\\ 
\tableline
A & 2400 & $-28$ & 38 & 64 & -  & -  & 01-08 14:52 $^{+6.6}_{-4.7}$ & $-14.9$  & 01-09 00:17 $^{+9.2}_{-6.9}$ & $-19.4$ & 01-09 18:22 $^{+7.6}_{-7.6}$ & $-28.1$\\
B & 2157 & $-25$ & 36 & 44 & 28 & 38 & 01-09 05:29 $^{+9.8}_{-5.7}$ &  $-0.4$  & 01-09 20:22 $^{+10.3}_{-6.6}$&   $+0.7$ & 01-11 03:13 $^{+26}_{-19}$   & $+4.7$ \\
B$'$&2157& $-25$ & 36 & 44 & -  & -  & 01-08 22:54 $^{+9.8}_{-6.8}$ & $-6.9$  & 01-09 11:42 $^{+11.0}_{-9.0}$ &   $-8.0$ & 01-10 12:10 $^{+22.2}_{-15.2}$ & $-10.3$\\
C & 2400 & $-28$ & 38 & 44 & -  & -  & 01-08 21:14 $^{+11.6}_{-9.0}$ & $-8.5$  & 01-09 08:29 $^{+16.0}_{-11.3}$& $-11.2$ & 01-10 07:21 $^{+8.6}_{-10.8}$ & $-15.2$\\
D & 2400 & $-28$ & 48 & 64 & -  & -  & 01-08 16:28 $^{+9.2}_{-5.9}$ & $-13.3$  & 01-09 02:25 $^{+12.3}_{-8.6}$& $-17.2$ & 01-09 17:47 $^{+8.5}_{-8.0}$  & $-28.7$\\
\tableline
\end{tabular}
\end{center}
\end{table*}

Since the GCS model provides the most reliable CME 3D reconstruction, it is anticipated that modeling CME propagation using GCS-derived parmeters, including tilt, should be more accurate when comparing to observations. To explore this, we completed another ensemble--B--simulation for this CME with parameters derived from GCS measurements and other analysis techniques described in Section \ref{kinematics} and shown in Figure \ref{Bpolar}, using an ellipsoid CME shape with a tilt.  This (and subsequent ensembles) use the same input WSA model synoptic map as Ensemble A.   Figure \ref{bnd} (right) shows a snapshot of the ENLIL model input inner boundary map for a time when the inserted tilted CME ellipsoid cross-section (ellipse) is largest (equal to the input CME minor and major axis half-widths), for one ensemble B member predicting a ``hit''. The ensemble B member shown has a speed of 2061 km\,s$^{-1}$, direction of 30$^{\circ}$ longitude, $-24^{\circ}$ latitude, 49$^{\circ}$ major axis half-width, and 31$^{\circ}$ minor axis half-width. For this model input, Figure \ref{fig:timB} shows a radial velocity contour plot in the same format as Figure \ref{fig:timA}. This simulation figure shows a small northeastern portion of the CME impacting Earth in contrast to Figure \ref{fig:timA}.

Figures \ref{stacksVenus}b, \ref{stacksEarth}b, and \ref{stacksMars}b show the modeled magnetic field, radial velocity, density, and temperature profiles at Venus, Earth, and Mars plotted as color traces for all 48 ensemble members.  Compared with ensemble A (see Figures \ref{stacksVenus}a-\ref{stacksMars}a), ensemble B results better match the {\it in-situ} observations at Earth and Venus, particularly the solar wind speed at Earth.  The modeled magnetic field is lower than observed at Venus and Earth because ENLIL is only modeling the effect of the pile-up of the plasma and magnetic field draping ahead of the CME and does not include the CME flux-rope. The modeling results show that the  probability of CME arrival is 38\% for Venus, 35\% for Earth, and 100\% for Mars.  

At Venus, the ensemble B mean arrival-time is on 9 January at 05:20 UT with individual arrival-times ranging from $+10/-6$ hours around the mean (indicated by the grey bar in \ref{stacksVenus}b), giving a prediction error of $-0.4$ hours (slight early prediction). At Earth, the ensemble mean arrival-time is on 9 January at 20:22 UT with individual arrival-times ranging from $+10/-7$ hours around the mean, giving a prediction error of +0.7 hours (slight late prediction). The standard deviation of the overall $K_{\rm P}$ forecast probability distribution for ensemble B is $\sigma$\,=\,0.8, and Figure \ref{kpAB}b shows that $\approx$\,94\% of the predictions fall between $K_{\rm P}$\,=\,3 to 5. Either using the average $K_{\rm P}$ prediction of 3 or the most likely $K_{\rm P}$ prediction of 4, the $K_{\rm P}$ prediction error is 0 or 1. 

At Mars, the ensemble mean arrival-time is on 11 January at 03:13 UT with individual arrival-times ranging from $+26/-19$ hours around the mean, giving a prediction error of +4.7 hours (late prediction).  The arrival-times at Mars have a large range with most arrivals near $\approx$\,22:00 UT on 10 January and some trailing late arrivals.  For ensemble B, the CME input speed is not the main distinguishing factor for arrival-times.  Instead, the earlier group of modeled arrivals correspond to CME inputs with major half-widths above $\approx$\,40$^{\circ}$, and the late group have major half-widths in the range of 32$^{\circ}$--39$^{\circ}$.  This group of modeled late arrivals suggests that CME major half-width is more likely to be above 40$^{\circ}$. Ensemble B results are summarized in Table \ref{tbl:summ} and show an overall improvement in the CME arrival-time and $K_{\rm P}$ prediction compared with A.

\begin{figure*}
\epsscale{1.0}
\plotone{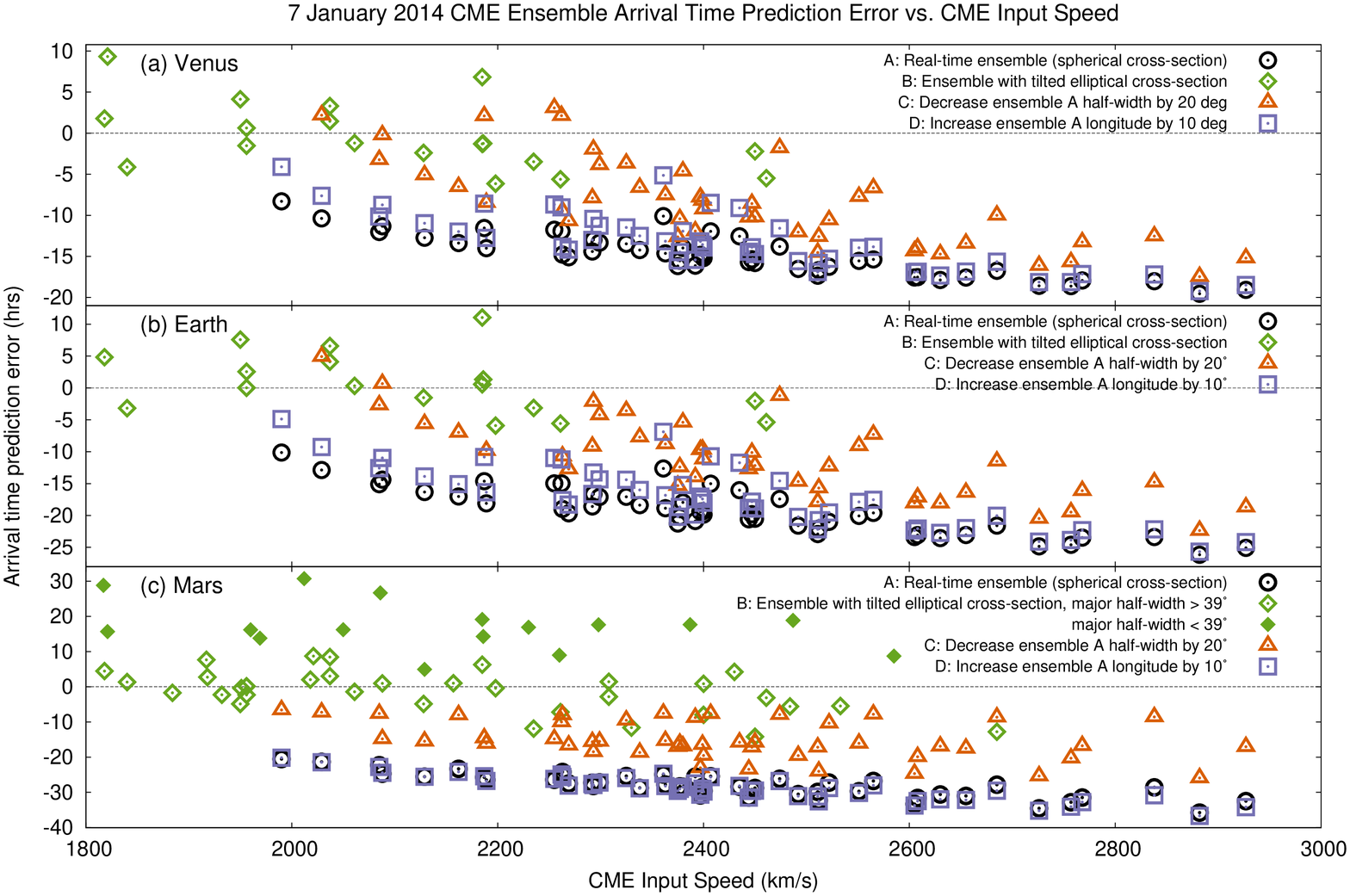}
\caption{\label{comparefig} Arrival-time prediction error for (a) Venus, (b) Earth, and (c) Mars plotted against the CME input speed for all ensemble members containing hits.  Ensemble A measurements and simulations were performed in real-time with a spherical CME shape, and are presented by black circles. Ensemble B (green diamonds) measurements and simulations were obtained with science level data and use an ellipsoid CME shape with a tilt from GCS measurements. Ensembles C (orange triangles) half-widths are reduced by 20$^{\circ}$ compared with ensemble A. Ensemble D (purple squares) longitudes are increased by 10$^{\circ}$ compared with ensemble A.}
\end{figure*}
\subsection{Comparing Ensemble Results}\label{compare}
In this section we compare ensemble A modeling results with B, which were described above in Sections \ref{ensa}--\ref{ensb} and listed in Table \ref{tbl:summ}, and consider other possible reasons for ensemble A's large prediction errors. Often, the CME width measurement is poorly constrained, and we note that the ellipse major half-widths of ensemble B members are on average $\approx$\,20$^{\circ}$ less than the spherical half-widths of ensemble A.  In order to better compare ensembles A and B, we constructed another ensemble\,--\,C\,--\,such that the members have the same CME parameters as ensemble A except the half-widths are reduced by 20$^{\circ}$. For ensemble C the prediction error for the average arrival-time becomes $-8.5$, $-11.2$, $-15.2$ hours for Venus, Earth, and Mars respectively (compared with $-14.9$, $-19.4$, $-28.1$ hours prediction error from ensemble A; see Table \ref{tbl:summ}).   Figure \ref{comparefig}a--c show the arrival-time prediction error for (a) Venus, (b) Earth, and (c) Mars plotted against the CME input speed for all of the ensemble members containing hits.  The arrival-time prediction improvement of ensemble C (orange triangles) compared with ensemble A (black circles) can be seen in this figure.  Compared with ensemble A, the ensemble C modeled arrival-time at Earth becomes later by $\approx$\,4 to 18 hours (8 hours on average), with the largest differences from CME leg arrivals, far from the nose (glancing arrival).  The prediction error improves by $\approx$\,2 to 15 hours (7 hours on average) for Venus and $\approx$\,8 to 20 hours (13 hours on average) for Mars. However, the prediction error improvement is not sufficient enough to explain the large early prediction error of ensemble A solely based on larger widths.  

Next, we consider the possibility of CME deflection or the accuracy of the CME longitude measurement. We observed that the median CME longitudes of ensemble A and B are 20$^{\circ}$--30$^{\circ}$ west of the flare source location longitude of 8$^{\circ}$. One possible cause for this difference was thought to be from the large coronal holes to the northeast of the active region (see Figure \ref{aia}a) that subsequently produced a high speed stream at Earth on 12 January reaching $\approx$\,850 km\,s$^{-1}$. However, \citet{mostl2015} suggest that the CME is channeled due to strong nearby active regions to the northeast and open coronal field to the west, rather than deflected by the coronal holes.  The GCS measurements did not show any deflection in the CME longitude in the measurement range of 2.1 to 18.5 R$_{\odot}$, which suggests the channeling occurred below this height.  Similarly, \citet{thalmann2015} show using nonlinear force free modeling that the coronal field overlying an active region influences CME speed and direction.  Since the CME input parameters making up these ensembles are based on coronagraph observations subsequent to the lower coronal channeling, and bearing in mind that CMEs in ENLIL are initialized at an inner boundary of 21.5 R$_{\odot}$, ensemble A arrival-time prediction errors cannot be explained by this longitudinal difference.  To illustrate this further, we constructed another ensemble\,--\,D\,--\,based on ensemble A, such that all of the CME input parameters remain unchanged except the longitude is increased by 10$^{\circ}$ for all members.   Ensemble D is denoted by purple squares in Figure \ref{comparefig} panels a--c. The prediction error for the average arrival-time becomes $-13.3$, $-17.2$, $-28.7$ hours for Venus, Earth, and Mars respectively (see Table \ref{tbl:summ}). Compared with ensemble A, the modeled arrival-time at Earth became later by 0.5 to 6 hours and 0.3 to 4 hours for Venus. For Mars, when increasing the longitude, the ensemble D modeled prediction error becomes worse on average by 1 hour compared with ensemble A. Therefore, increasing the longitude leads to a very minor prediction error improvement for Earth and Venus and is not nearly sufficient enough to explain the large early prediction error of ensemble A.   

The spread in arrival-times from the ensembles shown in Table \ref{tbl:summ} depends on a few factors which need to be considered when making comparisons.   The initial distribution of CME input parameters is the main controlling factor.  The smaller the spread in the CME parameter distribution (smaller error bars), the smaller the spread in arrival-times (higher confidence in the arrival-time).  Ensemble B has larger CME parameter error bars and therefore results in larger arrival-time spread compared with ensemble A.  The spread in arrival-times of glancing CME arrivals is expected to be larger (lower confidence in the arrival-time) than more central arrivals.  In these cases, a combination of the CME direction, width, and speed can produce a large spread in arrival-times reflecting impacts ranging from near CME center to the legs, as seen in ensembles C and D.

Compared with ensemble A, ensemble B's CME arrival-time and $K_{\rm P}$ predictions are greatly improved. The prediction error from the average ensemble A arrival-time was $-14.9$, $-19.4$, $-28.1$ hours compared with $-0.4$, $+0.7$, $+4.7$ hours for ensemble B for Venus, Earth, and Mars respectively.  This can be seen in Figure \ref{comparefig}a--c where the arrival-time prediction errors at Earth, Venus, and Mars are plotted against the CME input speed for all members containing hits in ensemble A (black circles) and B (green diamonds).  Comparing Figures \ref{stacksVenus}a--\ref{stacksMars}a and Figures \ref{stacksVenus}b--\ref{stacksMars}b the {\it in-situ} values and overall modeled arrival-time range is better captured by ensemble B. Ensemble A predicted a 78\% chance for $K_{\rm P}$ to reach between 6--8, compared with a 94\% chance for $K_{\rm P}$ to reach between 3--5 for Ensemble B (see Figures \ref{kpAB}a and \ref{kpAB}b).  As expected, because ensemble B better captures the {\it in-situ} observations, the probabilistic $K_{\rm P}$ prediction is closer to the observed value.  Ensemble B's $K_{\rm P}$ predictions also show improvement compared with ensemble D (88\% chance for $K_{\rm P}$\,=\,4--8, $\sigma$\,=\,1.6) and ensemble C (81\% chance for $K_{\rm P}$\,=\,3--7, $\sigma$\,=\,1.5).  We note that the most-likely and average predicted  $K_{\rm P}$ from ensemble C are similar to ensemble B, however the ensemble C prediction has less confidence (larger standard deviation). 

Next, we consider the background solar wind solution. Studies have shown that the CME arrival-times modeled by WSA--ENLIL+Cone are sensitive to the accuracy of the background solar wind solution \citep{lee2013,mays2015}. \citet{temmer2011} compared WSA--ENLIL modeled background solar wind with observations of CME kinematics in interplanetary space and found that the ambient solar wind is able to change the CME kinematics, for example slow CMEs become embedded in the background solar wind. However for this period, the model background solar wind prediction prior to CME arrival was close to the observed values of $\approx$\,350 km\,s$^{-1}$ (see Figure \ref{stacksEarth}). Nevertheless, we consider the effect of the GONG daily-updated synoptic magnetogram used by WSA to generate the ENLIL inner boundary conditions.  The original synoptic magnetogram used was for Carrington rotation number 2135 and 40$^{\circ}$ Carrington longitude  on 7 January 2014 at 18:04 UT.  When we change the synoptic magnetogram to a Carrington longitude of 50$^{\circ}$ (7 January at 01:04) and 30$^{\circ}$ Carrington longitude (8 January at 14:04 UT) the modeled arrival-times for ensemble A do not change by more than $\pm$1 hour for Venus and Earth.  For Mars, the modeled arrivals are shifted earlier by $\approx$\,2 and 6 hours for synoptic maps with Carrington longitude 30$^{\circ}$ and 50$^{\circ}$ respectively.  At Mars, the simulations from the different maps show a high speed stream on 8 January starting earlier in the day and reaching a higher speed, resulting in earlier CME arrival-times ($\approx$\,2 and 6 hours).

To isolate the effect of the ellipsoid CME shape and tilt used in ensemble B compared with the spheroid shape used in ensembles A, C, and D, we carried out modeling with ensemble B parameters using a spherical cross section (and no tilt). This ensemble is labelled B$'$ in Table \ref{tbl:summ}. Modeled CME arrival-times and $K_{\rm P}$ predictions for this ensemble are very similar to ensemble C (ensemble A half-width was decreased by 20$^{\circ}$) with a small average arrival-time improvement of 1, 3, and 5 hours at Venus, Earth, and Mars respectively.   Finally, to quantify the effect of prior CMEs on the 7 January CME propagation we performed additional simulations that included the prior 4 January and 6 January CMEs (discussed in Section \ref{obs}).  When the prior CMEs are included, the 7 January CME is modeled to arrive $\approx$\,1, 2, and 3 hours earlier at Venus, Earth, and Mars respectively than when they are not included (increasing the prediction error).  Taken together, these modeling results suggest that the CME orientation and directionality with respect to observatories at Venus, Earth, and Mars may play an important role in understanding the propagation of this CME.

\section{Summary of Results}\label{disc}
We presented a series of ensemble modeling results for the interplanetary propagation of the 7 January 2014 CME and arrival at Venus, Earth, and Mars with the WSA--ENLIL+Cone model.  The 7 January CME was chosen for this modeling exercise because its arrival was much later than most of the predicted arrival-times on the {\it CME Scoreboard}, and was far outside the associated range of errors for the various methods used.  There are many potential reasons for the inaccuracy of the WSA--ENLIL+Cone prediction; arrival-times depend on the accurate specification of early CME kinematics and geometry, and the model's ability to reproduce background solar wind conditions.  For example, a few days later after the CME arrival (12 January) a high speed stream reaching $\approx$\,850 km\,s$^{-1}$ was observed.  Considering the large difference ($\approx$\,37$^{\circ}$) in propagation direction relative to the source region longitude, it was thought at first that the coronal holes could have deflected the CME from its initial path. However, \citet{mostl2015} suggest that the CME is channeled due to strong nearby active regions to the northeast and open coronal field to the west rather than deflected by the coronal holes.  Also, there was a delay of more than a day between the {\it in-situ} CME and high speed stream signatures and the interstitial measurements indicated standard slow wind with no signs of compression or interaction.  Finally, GCS measurements did not show any deflection in the CME longitude from 2.1 to 18.5 R$_{\odot}$, so the channeling happens in the low corona (see {\it e.g.} \citet{thalmann2015}), which indicates that the CME arrival delay is unlikely to be due to change in CME longitude in the outer corona.

The real-time WSA--ENLIL+Cone results from ensemble A (using default model settings) were systematically compared with post-event ensemble simulations each testing different effects. We examined the impact of CME shape and tilt (ensemble B), CME half-width (ensemble C), CME longitude (ensemble D), changing the input magnetogram, and including prior CMEs.   The prediction error from the average ensemble A arrival-time was $-14.9$, $-19.4$, $-28.1$ hours compared with $-0.4$, $+0.7$, $+4.7$ hours for ensemble B for Venus, Earth, and Mars respectively; ensemble B shows a clear improvement.  Because ensemble B better captures the {\it in-situ} observations, the probabilistic $K_{\rm P}$ prediction for ensemble B (94\% chance for $K_{\rm P}$ to reach between 3--5) is closer to the observed value compared with ensemble A (78\% chance for $K_{\rm P}$ to reach between 6--8). ENLIL models the effect of the pile-up of the plasma and magnetic field draping ahead of the CME and does not include the CME flux-rope and a magnetic cloud was not observed {\it in-situ}; therefore it was not possible to investigate the effect of the intrinsic CME magnetic field.  However, the different effects considered the simulations show that using a tilted ellipsoid CME shape with major and minor axis half-widths would have improved the initial real-time prediction to better reflect the observed {\it in-situ} signatures at Venus, Earth, and Mars and the geomagnetic storm strength.  

The modeling results suggest that CME orientation and directionality with respect to observatories at Venus, Earth, and Mars play an important role in understanding the propagation of this CME.  This suggests that CME orientation and shape should be taken into consideration when modeling the propagation of glancing CME arrivals, whether due to their asymmetric shape in latitude and longitude, or tilt with respect to the solar equator.  The 7 January 2014 event illustrates that this effect plays a particularly important role for CMEs which have a glancing arrival at the location of interest.  Many CME propagation models do not currently take into account the full 3D geometry of the CME, such as tilt, or propagation direction out of the ecliptic plane (2D models). For 2D models, the cross section of a GCS reconstructed CME along the ecliptic plane could be used to improve such modeling results. It is important to note that GCS model fits of CMEs are not possible without multiple viewpoint imaging such as those from {\it STEREO}/SECCHI.    Finally, when the modeled CME time of arrival is improved this  leads to an improvement in the modeled radial velocity of arrival when comparing to {\it in-situ} observations, which in turn can lead to better ENLIL-based $K_{\rm P}$ predictions. For example, \citet{savani2015b} show how better estimating which part of the CME structure is sampled by an observer during its transit is important for improving $K_{\rm P}$ predictions.


\acknowledgments

We gratefully acknowledge contributions from the model developers and participants of the Scoreboard (\href{http://kauai.ccmc.gsfc.nasa.gov/CMEscoreboard}{{\sf kauai.ccmc.gsfc.nasa.gov/CMEscoreboard}}).   M.L.M. thanks T. Nieves-Chinchilla, I.G. Richardson, J.G. Luhmann, and N. Thakur for helpful discussions. M.L.M. acknowledges the support of NASA LWS grant NNX15AB80G. L.K. Jian acknowledges the support of NSF grants AGS 1242798, 1259549, and 1321493. R.C.C. acknowledges the support of NASA contract S-136361-Y to NRL. D. Odstrcil acknowledges the support of NASA LWS-SC NNX13AI96G. C.M. and M.T. acknowledge the Austrian Science Fund (FWF): P26174-N27 and V195-N16. The presented work has received funding from the European Union Seventh Framework Programme (FP7/2007–2013) under grant agreement No. 606692 [HELCATS].The {\it ACE}, {\it Wind}, and OMNI solar wind plasma and magnetic field data were obtained at NASA's CDAWeb (\href{http://cdaweb.gsfc.nasa.gov}{{\sf cdaweb.gsfc.nasa.gov}} and OMNIWeb). The final definitive $K_{\rm P}$ indices were obtained from the Helmholtz Center Potsdam GFZ German Research Centre for Geosciences. The PFSS modeled magnetic field lines were traced using SolarSoft's PFSS package. {\it SOHO} is a mission of international cooperation between the European Space Agency and NASA. Some figure colors are based on \href{http://www.colorbrewer.org}{{\sf ColorBrewer.org}}.


\bibliographystyle{apj}
\bibliography{jan7ref}  

\begin{thebibliography}{}
\expandafter\ifx\csname natexlab\endcsname\relax\def\natexlab#1{#1}\fi

\bibitem[{{Arge} {et~al.}(2004){Arge}, {Luhmann}, {Odstr{\v c}il}, {Schrijver},
  \& {Li}}]{arge2004}
{Arge}, C.~N., {Luhmann}, J.~G., {Odstr{\v c}il}, D., {Schrijver}, C.~J., \&
  {Li}, Y. 2004, \jastp, 66, 1295

\bibitem[{{Arge} \& {Pizzo}(2000)}]{arge2000}
{Arge}, C.~N., \& {Pizzo}, V.~J. 2000, \jgr, 105, 10465

\bibitem[{{Bothmer} \& {Schwenn}(1994)}]{bothmer1994}
{Bothmer}, V., \& {Schwenn}, R. 1994, \ssr, 70, 215

\bibitem[{{Bothmer} \& {Schwenn}(1998)}]{bothmer1998}
---. 1998, Annales Geophysicae, 16, 1

\bibitem[{{Brueckner} {et~al.}(1995){Brueckner}, {Howard}, {Koomen},
  {Korendyke}, {Michels}, {Moses}, {Socker}, {Dere}, {Lamy}, {Llebaria},
  {Bout}, {Schwenn}, {Simnett}, {Bedford}, \& {Eyles}}]{lasco}
{Brueckner}, G.~E., {Howard}, R.~A., {Koomen}, M.~J., {et~al.} 1995, \solphys,
  162, 357

\bibitem[{{Colaninno} {et~al.}(2013){Colaninno}, {Vourlidas}, \&
  {Wu}}]{colaninno2013}
{Colaninno}, R.~C., {Vourlidas}, A., \& {Wu}, C.~C. 2013, \jgr, 118, 6866

\bibitem[{{Domingo} {et~al.}(1995){Domingo}, {Fleck}, \& {Poland}}]{soho}
{Domingo}, V., {Fleck}, B., \& {Poland}, A.~I. 1995, \solphys, 162, 1

\bibitem[{{Emmons} {et~al.}(2013){Emmons}, {Acebal}, {Pulkkinen},
  {Taktakishvili}, {MacNeice}, \& {Odstr{\v c}il}}]{emmons2013}
{Emmons}, D., {Acebal}, A., {Pulkkinen}, A., {et~al.} 2013, Space Weather, 11,
  95

\bibitem[{{Gopalswamy} {et~al.}(2009){Gopalswamy}, {Yashiro}, {Michalek},
  {Stenborg}, {Vourlidas}, {Freeland}, \& {Howard}}]{gopalswamy2009}
{Gopalswamy}, N., {Yashiro}, S., {Michalek}, G., {et~al.} 2009, Earth Moon and
  Planets, 104, 295

\bibitem[{{Hapgood}(1992)}]{hapgood1992}
{Hapgood}, M.~A. 1992, \planss, 40, 711

\bibitem[{{Harvey} {et~al.}(1996){Harvey}, {Hill}, {Hubbard}, {Kennedy},
  {Leibacher}, {Pintar}, {Gilman}, {Noyes}, {Title}, {Toomre}, {Ulrich},
  {Bhatnagar}, {Kennewell}, {Marquette}, {Patron}, {Saa}, \&
  {Yasukawa}}]{harvey1996}
{Harvey}, J.~W., {Hill}, F., {Hubbard}, R.~P., {et~al.} 1996, Science, 272,
  1284

\bibitem[{{Howard} {et~al.}(2008){Howard}, {Moses}, {Vourlidas}, {Newmark},
  {Socker}, {Plunkett}, \& {\it et al.}}]{secchi}
{Howard}, R.~A., {Moses}, J.~D., {Vourlidas}, A., {et~al.} 2008, \ssr, 136, 67

\bibitem[{{Hundhausen} {et~al.}(1994){Hundhausen}, {Burkepile}, \&
  {St.~Cyr}}]{hundhausen1994}
{Hundhausen}, A.~J., {Burkepile}, J.~T., \& {St.~Cyr}, O.~C. 1994, \jgr, 99,
  6543

\bibitem[{{Jian} {et~al.}(2006){Jian}, {Russell}, {Luhmann}, \&
  {Skoug}}]{jian2006}
{Jian}, L., {Russell}, C.~T., {Luhmann}, J.~G., \& {Skoug}, R.~M. 2006,
  \solphys, 239, 393

\bibitem[{{Kaiser} {et~al.}(2008){Kaiser}, {Kucera}, {Davila}, {St.~Cyr},
  {Guhathakurta}, \& {Christian}}]{stereo}
{Kaiser}, M.~L., {Kucera}, T.~A., {Davila}, J.~M., {et~al.} 2008, \ssr, 136, 5

\bibitem[{{Kim} {et~al.}(2008){Kim}, {Cho}, {Kim}, {Park}, {Moon}, {Yi}, {Lee},
  {Wang}, {Song}, \& {Dryer}}]{kim2008}
{Kim}, R.-S., {Cho}, K.-S., {Kim}, K.-H., {et~al.} 2008, \apj, 677, 1378

\bibitem[{{Kwon} {et~al.}(2015){Kwon}, {Zhang}, \& {Vourlidas}}]{kwon2015}
{Kwon}, R.-Y., {Zhang}, J., \& {Vourlidas}, A. 2015, \apjl, 799, L29

\bibitem[{{Lee} {et~al.}(2013){Lee}, {Arge}, {Odstr{\v c}il}, {Millward},
  {Pizzo}, {Quinn}, \& {Henney}}]{lee2013}
{Lee}, C.~O., {Arge}, C.~N., {Odstr{\v c}il}, D., {et~al.} 2013, \solphys, 285,
  349

\bibitem[{{Lemen} {et~al.}(2012){Lemen}, {Title}, {Akin}, {Boerner}, {Chou},
  {Drake}, {Duncan}, {Edwards}, {Friedlaender}, {Heyman}, {Hurlburt}, {Katz},
  {Kushner}, {Levay}, {Lindgren}, {Mathur}, {McFeaters}, {Mitchell}, {Rehse},
  {Schrijver}, {Springer}, {Stern}, {Tarbell}, {Wuelser}, {Wolfson}, {Yanari},
  {Bookbinder}, {Cheimets}, {Caldwell}, {Deluca}, {Gates}, {Golub}, {Park},
  {Podgorski}, {Bush}, {Scherrer}, {Gummin}, {Smith}, {Auker}, {Jerram},
  {Pool}, {Soufli}, {Windt}, {Beardsley}, {Clapp}, {Lang}, \&
  {Waltham}}]{lemen2012}
{Lemen}, J.~R., {Title}, A.~M., {Akin}, D.~J., {et~al.} 2012, \solphys, 275, 17

\bibitem[{{Liu}(2008)}]{liu2008}
{Liu}, Y. 2008, \apjl, 679, L151

\bibitem[{{Liu} {et~al.}(2010){Liu}, {Thernisien}, {Luhmann}, {Vourlidas},
  {Davies}, {Lin}, \& {Bale}}]{liu2010b}
{Liu}, Y., {Thernisien}, A., {Luhmann}, J.~G., {et~al.} 2010, \apj, 722, 1762

\bibitem[{{Mays} {et~al.}(2015){Mays}, {Taktakishvili}, {Pulkkinen},
  {MacNeice}, {Rast{\"a}tter}, {Odstrcil}, {Jian}, {Richardson}, {LaSota},
  {Zheng}, \& {Kuznetsova}}]{mays2015}
{Mays}, M.~L., {Taktakishvili}, A., {Pulkkinen}, A., {et~al.} 2015, \solphys,
  290, doi:10.1007/s11207-015-0692-1

\bibitem[{{Mierla} {et~al.}(2010){Mierla}, {Inhester}, {Antunes}, {Boursier},
  {Byrne}, {Colaninno}, {Davila}, {de Koning}, {Gallagher}, {Gissot}, {Howard},
  {Howard}, {Kramar}, {Lamy}, {Liewer}, {Maloney}, {Marqu{\'e}}, {McAteer},
  {Moran}, {Rodriguez}, {Srivastava}, {St.~Cyr}, {Stenborg}, {Temmer},
  {Thernisien}, {Vourlidas}, {West}, {Wood}, \& {Zhukov}}]{mierla2010}
{Mierla}, M., {Inhester}, B., {Antunes}, A., {et~al.} 2010, Annales
  Geophysicae, 28, 203

\bibitem[{{Moon} {et~al.}(2005){Moon}, {Cho}, {Dryer}, {Kim}, {Bong}, {Chae},
  \& {Park}}]{moon2005}
{Moon}, Y.-J., {Cho}, K.-S., {Dryer}, M., {et~al.} 2005, \apj, 624, 414

\bibitem[{{M{\"o}stl} {et~al.}(2014){M{\"o}stl}, {Amla}, {Hall}, {Liewer}, {De
  Jong}, {Colaninno}, {Veronig}, {Rollett}, {Temmer}, {Peinhart}, {Davies},
  {Lugaz}, {Liu}, {Farrugia}, {Luhmann}, {Vr{\v s}nak}, {Harrison}, \&
  {Galvin}}]{mostl2014}
{M{\"o}stl}, C., {Amla}, K., {Hall}, J.~R., {et~al.} 2014, \apj, 787, 119

\bibitem[{{M\"{o}stl} {et~al.}(2015){M\"{o}stl}, {Rollett}, {Frahm}, {Liu},
  {Long}, {Colaninno}, {Reiss}, {Temmer}, {Farrugia}, {Posner}, {Dumbovi{\'c}},
  {Janvier}, {D\'{e}moulin}, {Boakes}, {Devos}, {Kraaikamp}, {Mays}, \& {r{\v
  s}nak}}]{mostl2015}
{M\"{o}stl}, C., {Rollett}, T., {Frahm}, R., {et~al.} 2015, Nature
  Communications, 6, 7135

\bibitem[{{Newell} {et~al.}(2007){Newell}, {Sotirelis}, {Liou}, {Meng}, \&
  {Rich}}]{newell2007}
{Newell}, P.~T., {Sotirelis}, T., {Liou}, K., {Meng}, C.-I., \& {Rich}, F.~J.
  2007, \jgr, 112, 1206

\bibitem[{{Odstr{\v c}il}(2003)}]{odstrcil2003}
{Odstr{\v c}il}, D. 2003, \adv, 32, 497

\bibitem[{{Odstr{\v c}il} \& {Pizzo}(1999{\natexlab{a}})}]{odstrcil1999_1}
{Odstr{\v c}il}, D., \& {Pizzo}, V.~J. 1999{\natexlab{a}}, \jgr, 104, 483

\bibitem[{{Odstr{\v c}il} \& {Pizzo}(1999{\natexlab{b}})}]{odstrcil1999_2}
---. 1999{\natexlab{b}}, \jgr, 104, 493

\bibitem[{{Odstr{\v c}il} {et~al.}(2004){Odstr{\v c}il}, {Riley}, \&
  {Zhao}}]{odstrcil2004}
{Odstr{\v c}il}, D., {Riley}, P., \& {Zhao}, X.~P. 2004, \jgr, 109, 2116

\bibitem[{{Odstr{\v c}il} {et~al.}(1996){Odstr{\v c}il}, {Smith}, \&
  {Dryer}}]{ods1996}
{Odstr{\v c}il}, D., {Smith}, Z., \& {Dryer}, M. 1996, \grl, 23, 2521

\bibitem[{{Pesnell} {et~al.}(2012){Pesnell}, {Thompson}, \&
  {Chamberlin}}]{pesnell2012}
{Pesnell}, W.~D., {Thompson}, B.~J., \& {Chamberlin}, P.~C. 2012, \solphys,
  275, 3

\bibitem[{{Savani} {et~al.}(2013){Savani}, {Vourlidas}, {Pulkkinen},
  {Nieves-Chinchilla}, {Lavraud}, \& {Owens}}]{savani2013}
{Savani}, N.~P., {Vourlidas}, A., {Pulkkinen}, A., {et~al.} 2013, Space
  Weather, 11, 245

\bibitem[{{Savani} {et~al.}(submitted, 2015){Savani}, {Vourlidas},
  {Richardson}, {Szabo}, {Thompson}, {Pulkkinen}, {Mays}, \&
  {Evans}}]{savani2015b}
{Savani}, N.~P., {Vourlidas}, A., {Richardson}, I.~G., {et~al.} submitted,
  2015, Space Weather

\bibitem[{{Savani} {et~al.}(2015){Savani}, {Vourlidas}, {Szabo}, {Mays},
  {Thompson}, {Richardson}, {Evans}, {Pulkkinen}, \&
  {Nieves-Chinchilla}}]{savani2015a}
{Savani}, N.~P., {Vourlidas}, A., {Szabo}, A., {et~al.} 2015, Space Weather,
  13, doi:10.1002/2015SW001171

\bibitem[{{Scherrer} {et~al.}(2012){Scherrer}, {Schou}, {Bush}, {Kosovichev},
  {Bogart}, {Hoeksema}, {Liu}, {Duvall}, {Zhao}, {Title}, {Schrijver},
  {Tarbell}, \& {Tomczyk}}]{scherrer2012}
{Scherrer}, P.~H., {Schou}, J., {Bush}, R.~I., {et~al.} 2012, \solphys, 275,
  207

\bibitem[{{Schrijver} {et~al.}(2006){Schrijver}, {De Rosa}, {Metcalf}, {Liu},
  {McTiernan}, {R{\'e}gnier}, {Valori}, {Wheatland}, \&
  {Wiegelmann}}]{schrijver2006}
{Schrijver}, C.~J., {De Rosa}, M.~L., {Metcalf}, T.~R., {et~al.} 2006,
  \solphys, 235, 161

\bibitem[{{Schwenn} {et~al.}(2005){Schwenn}, {dal Lago}, {Huttunen}, \&
  {Gonzalez}}]{schwenn2005}
{Schwenn}, R., {dal Lago}, A., {Huttunen}, E., \& {Gonzalez}, W.~D. 2005,
  Annales Geophysicae, 23, 1033

\bibitem[{{Shen} {et~al.}(2014){Shen}, {Wang}, {Pan}, {Miao}, {Ye}, \&
  {Wang}}]{shen2014}
{Shen}, C., {Wang}, Y., {Pan}, Z., {et~al.} 2014, \jgr, 119, 5107

\bibitem[{{Shi} {et~al.}(2015){Shi}, {Wang}, {Wan}, {Cheng}, {Ding}, \&
  {Zhang}}]{shi2015}
{Shi}, T., {Wang}, Y., {Wan}, L., {et~al.} 2015, {arXiv e-prints},
  arXiv:1505.00884

\bibitem[{{Stone} {et~al.}(1998){Stone}, {Frandsen}, {Mewaldt}, {Christian},
  {Margolies}, {Ormes}, \& {Snow}}]{ace}
{Stone}, E.~C., {Frandsen}, A.~M., {Mewaldt}, R.~A., {et~al.} 1998, \ssr, 86, 1

\bibitem[{{Svedhem} {et~al.}(2009){Svedhem}, {Titov}, {Taylor}, \&
  {Witasse}}]{svedhem2009}
{Svedhem}, H., {Titov}, D., {Taylor}, F., \& {Witasse}, O. 2009, Journal of
  Geophysical Research (Planets), 114, 0

\bibitem[{{Temmer} {et~al.}(2011){Temmer}, {Rollett}, {M{\"o}stl}, {Veronig},
  {Vr{\v s}nak}, \& {Odstr{\v c}il}}]{temmer2011}
{Temmer}, M., {Rollett}, T., {M{\"o}stl}, C., {et~al.} 2011, \apj, 743, 101

\bibitem[{{Thakur} {et~al.}(2014){Thakur}, {Gopalswamy}, {Xie},
  {M{\"a}kel{\"a}}, {Yashiro}, {Akiyama}, \& {Davila}}]{thakur2014}
{Thakur}, N., {Gopalswamy}, N., {Xie}, H., {et~al.} 2014, \apjl, 790, L13

\bibitem[{{Thalmann} {et~al.}(2015){Thalmann}, {Su}, {Temmer}, \&
  {Veronig}}]{thalmann2015}
{Thalmann}, J.~K., {Su}, Y., {Temmer}, M., \& {Veronig}, A.~M. 2015, \apjl,
  801, L23

\bibitem[{{Thernisien} {et~al.}(2009){Thernisien}, {Vourlidas}, \&
  {Howard}}]{thernisien2009}
{Thernisien}, A., {Vourlidas}, A., \& {Howard}, R.~A. 2009, \solphys, 256, 111

\bibitem[{{Thernisien} {et~al.}(2006){Thernisien}, {Howard}, \&
  {Vourlidas}}]{thernisien2006}
{Thernisien}, A.~F.~R., {Howard}, R.~A., \& {Vourlidas}, A. 2006, \apj, 652,
  763

\bibitem[{{Thompson} \& {Young}({submitted, 2015})}]{thompson2015b}
{Thompson}, B.~J., \& {Young}, A. {submitted, 2015}, \apj

\bibitem[{{Thompson} {et~al.}(2015){Thompson}, {DeRosa}, {Fisher}, {Krista},
  {Kwon}, {Mason}, {Mays}, {Nitta}, {Webb}, \& {West}}]{thompson2015a}
{Thompson}, B.~J., {DeRosa}, M.~L., {Fisher}, R.~R., {et~al.} 2015, in AAS/AGU
  Triennial Earth-Sun Summit, Vol.~1, AAS/AGU Triennial Earth-Sun Summit, 21201

\bibitem[{{Thompson}(2006)}]{thompson2006}
{Thompson}, W.~T. 2006, \aap, 449, 791

\bibitem[{{T{\"o}r{\"o}k} \& {Kliem}(2005)}]{torok2005}
{T{\"o}r{\"o}k}, T., \& {Kliem}, B. 2005, \apjl, 630, L97

\bibitem[{{Vourlidas} {et~al.}(2013){Vourlidas}, {Lynch}, {Howard}, \&
  {Li}}]{vourlidas2013}
{Vourlidas}, A., {Lynch}, B.~J., {Howard}, R.~A., \& {Li}, Y. 2013, \solphys,
  284, 179

\bibitem[{{Vourlidas} {et~al.}(2003){Vourlidas}, {Wu}, {Wang}, {Subramanian},
  \& {Howard}}]{vourlidas2003}
{Vourlidas}, A., {Wu}, S.~T., {Wang}, A.~H., {Subramanian}, P., \& {Howard},
  R.~A. 2003, \apj, 598, 1392

\bibitem[{{Vr{\v s}nak} {et~al.}(2004){Vr{\v s}nak}, {Ru{\v z}djak}, {Sudar},
  \& {Gopalswamy}}]{vrsnak2004}
{Vr{\v s}nak}, B., {Ru{\v z}djak}, D., {Sudar}, D., \& {Gopalswamy}, N. 2004,
  \aap, 423, 717

\bibitem[{{Vr{\v s}nak} {et~al.}(2013){Vr{\v s}nak}, {{\v Z}ic}, {Vrbanec},
  {Temmer}, {Rollett}, {M{\"o}stl}, {Veronig}, {{\v C}alogovi{\'c}},
  {Dumbovi{\'c}}, {Luli{\'c}}, {Moon}, \& {Shanmugaraju}}]{vrsnak2013}
{Vr{\v s}nak}, B., {{\v Z}ic}, T., {Vrbanec}, D., {et~al.} 2013, \solphys, 285,
  295

\bibitem[{{Wang} {et~al.}(2002){Wang}, {Ye}, {Wang}, {Zhou}, \&
  {Wang}}]{wang2002}
{Wang}, Y.~M., {Ye}, P.~Z., {Wang}, S., {Zhou}, G.~P., \& {Wang}, J.~X. 2002,
  \jgr, 107, 1340

\bibitem[{{Xie} {et~al.}(2004){Xie}, {Ofman}, \& {Lawrence}}]{xie2004}
{Xie}, H., {Ofman}, L., \& {Lawrence}, G. 2004, \jgr, 109, 3109

\bibitem[{{Zhao} {et~al.}(2002){Zhao}, {Plunkett}, \& {Liu}}]{zhao2002}
{Zhao}, X.~P., {Plunkett}, S.~P., \& {Liu}, W. 2002, \jgr, 107, 1223

\end{thebibliography}

\end{document}